\documentclass[journal]{new-aiaa}
\usepackage[utf8]{inputenc}
\usepackage{graphicx}
\usepackage[version=4]{mhchem}
\usepackage{siunitx}
\usepackage{longtable,tabularx}
\usepackage{pdfsync} 
\usepackage{subcaption} 	
\usepackage{amsmath} 		
\usepackage{algorithmic}  	
\usepackage{algorithm}    	
\usepackage{mathtools}    	
\usepackage{float}        	
\usepackage{hyperref}     	
\usepackage{enumitem}     	
\usepackage{booktabs}     	
\usepackage{multirow}     	
\usepackage{makecell}     	
\usepackage{bm}           	
\newtheorem{remark}{Remark}[section]
\newtheorem{theorem}{Theorem}



\title{Sparse Kalman Identification for Partially Observable Systems via Adaptive Bayesian Learning}

\author{Jilan Mei \footnote{Ph.D. Candidate, School of Astronautics, zb2515202@buaa.edu.cn.} 
and
Tengjie Zheng \footnote{Ph.D. Candidate, School of Astronautics, ZhengTengjie@buaa.edu.cn.}
and
Lin Cheng \footnote{\textbf{Corresponding Author}, Associate Professor, School of Astronautics, chenglin5580@buaa.edu.cn.}
and
Shengping Gong \footnote{Professor, School of Astronautics, gongsp@buaa.edu.cn.}
and
Xu Huang \footnote{Associate Professor, School of Astronautics, xunudt@126.com.} 
}

\affil{School of Astronautics, Beihang University, Beijing, 102206, China}
\affil{State Key Laboratory of High-Efficiency Reusable Aerospace Transportation Technology, Beijing, 102206, China}

\begin{document}

\maketitle

\begin{abstract}
Sparse dynamics identification is an essential tool for discovering interpretable physical models and enabling efficient control in engineering systems. However, existing methods rely on batch learning with full historical data, limiting their applicability to real-time scenarios involving sequential and partially observable data. To overcome this limitation, this paper proposes an online Sparse Kalman Identification (SKI) method by integrating the Augmented Kalman Filter (AKF) and Automatic Relevance Determination (ARD). The main contributions are: (1) a theoretically grounded Bayesian sparsification scheme that is seamlessly integrated into the AKF framework and adapted to sequentially collected data in online scenarios; (2) an update mechanism that adapts the Kalman posterior to reflect the updated selection of the basis functions that define the model structure; (3) an explicit gradient-descent formulation that enhances computational efficiency. Consequently, the SKI method achieves accurate model structure selection with millisecond-level efficiency and higher identification accuracy, as demonstrated by extensive simulations and real-world experiments (showing an 84.21\% improvement in accuracy over the baseline AKF).
\end{abstract}

\section{Introduction}
\label{sec:introduction}

\lettrine{D}{ynamics} identification is a fundamental methodology for uncovering the underlying physical laws and is essential for control and decision-making in engineering systems~\cite{DMS1, DMS2, DMS3}. The growing complexity of modern systems has brought increasing attention to data-driven methods for dynamics identification. These methods offer flexibility and capability to model highly nonlinear behaviors without explicit first-principles knowledge, and have been widely applied across diverse domains, including fluid dynamics~\cite{PIT, FLU, SINDY, Rosafalco2025}, hypersonic aerodynamics~\cite{HP1, HP2, HP3}, adaptive control of drones~\cite{UAV1, UAV2, UAV3}, and even neurodynamics in brain–machine interfaces~\cite{BMI1, BMI2}. Motivated by these advances, this work focuses on addressing the corresponding challenges of data-driven dynamics identification.

\begin{figure*}[htbp]
  \centering
  \begin{subfigure}[b]{0.32\textwidth}
    \centering
    \includegraphics[width=\textwidth]{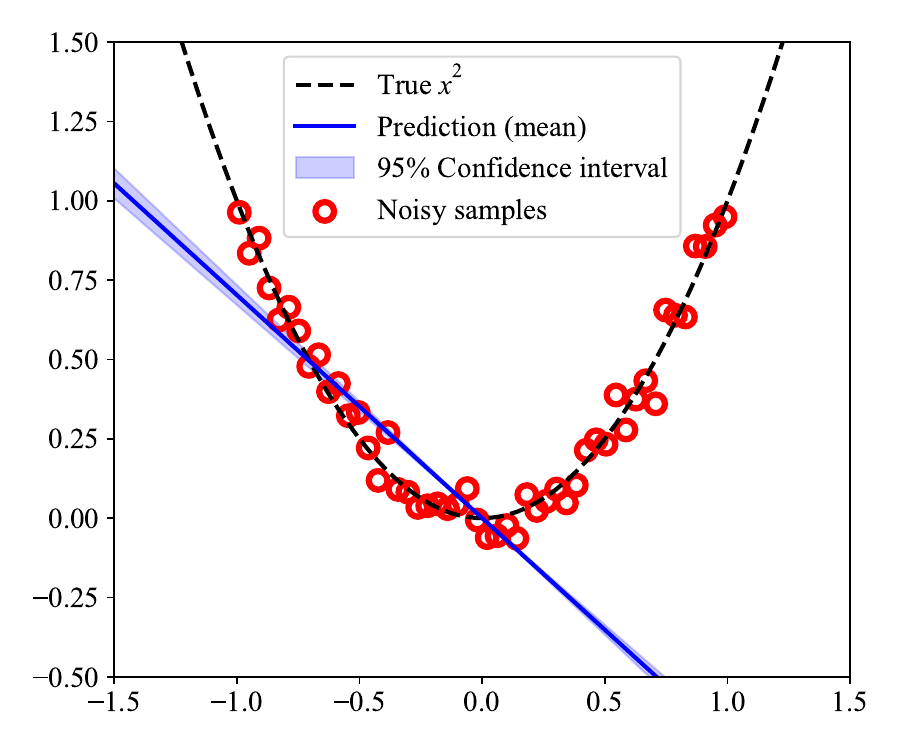}
    \caption{Underfitting}
    \label{fig:1a}
  \end{subfigure}
  \hfill
  \begin{subfigure}[b]{0.32\textwidth}
    \centering
    \includegraphics[width=\textwidth]{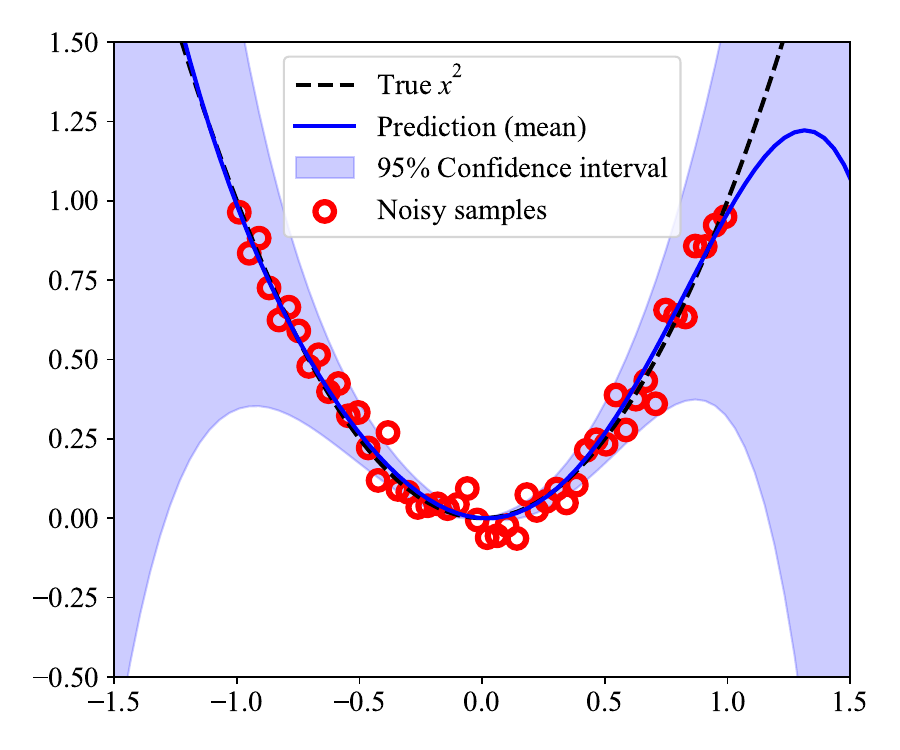}
    \caption{Overfitting}
    \label{fig:1b}
  \end{subfigure}
  \hfill
  \begin{subfigure}[b]{0.32\textwidth}
    \centering
    \includegraphics[width=\textwidth]{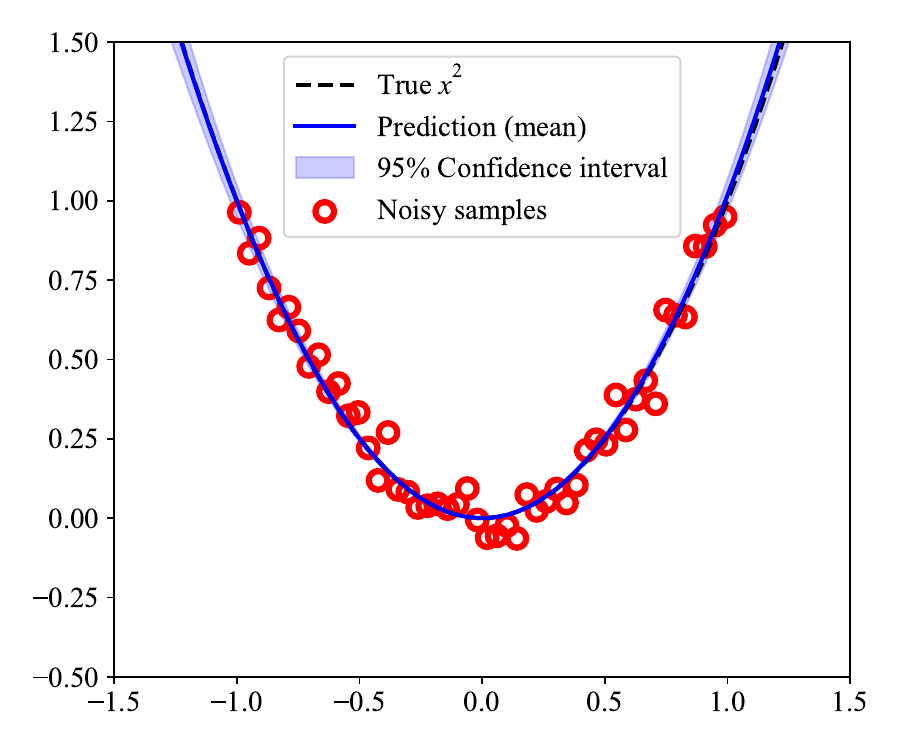}
    \caption{Sparse Identification}
    \label{fig:1c}
  \end{subfigure}
  \caption{Illustration of polynomial fitting to a quadratic function: (a) underfitting due to an overly simple basis, (b) overfitting resulting from an excessively complex basis, and (c) sparse identification yielding an accurate and parsimonious model.}
  \label{fig:1}
\end{figure*}

Data-driven identification methods can be broadly categorized into two classes: non-parametric methods and parametric methods. Non-parametric methods, free from prior structural knowledge, have demonstrated accurate and flexible capabilities in modeling complex systems. Typical approaches include Deep Neural Network (DNN), Radial Basis Function Neural Network (RBFNN), and Gaussian Processes (GP). The DNN methods describe the unknown mappings of the dynamical equations through deep neural networks, enabling the capture of highly nonlinear dynamics~\cite{CLNN, CHEN01081992, ERNN, Ji2024}. Similarly, the RBFNN methods use a single hidden layer of localized radial basis functions to describe the unknown mappings, enabling fast training and effective approximation of nonlinear dynamics~\cite{RBFNN1, RBFNN2}. Moreover, the GP methods model unknown dynamical mappings as distributions over functions, enabling both prediction and uncertainty quantification, which is crucial for adaptive control~\cite{gp_1, ZHENG2025110049, wingrock2GP}. Despite these advantages, non-parametric methods often require substantial training time and computational resources, and their limited interpretability restricts their applicability in safety-critical scenarios. In contrast, parametric identification methods integrate prior structural knowledge with data to determine the system dynamics. Typically, these methods represent unknown dynamical mappings as a linear combination of basis functions, with basis functions selected based on expert knowledge, transforming the dynamics identification into a data-driven parameter estimation problem. When the dynamics can be constructed as a regressive problem, Recursive Least Squares (RLS) methods can be directly applied for efficient parameter estimation. When the system is partially observable, the Augmented Kalman Filter (AKF) framework augments the weight parameters to state space, enabling joint estimation with the full system states~\cite{KF, AKF, Zhang2024}. When time delays are present in the system, frequency-domain methods offer advantages by providing robust parameter estimation through the analysis of system frequency responses~\cite{FTR1, FTR2}. Despite these advantages, a fundamental challenge arises in the selection of the basis functions: a basis set that is overly simplistic may result in insufficient accuracy (underfitting), whereas an excessively complex basis increases the risk of overfitting. Therefore, the judicious selection of basis functions is crucial to simultaneously ensure both model accuracy and generalization capability, as illustrated in Fig.~\ref{fig:1}.


To address this trade-off in the selection of basis functions, sparse identification has emerged as an important approach. By driving the weight parameters of irrelevant basis functions to zero, these methods remove redundant functions from the candidate set, enabling a dynamic balance between precision and parsimony. One classical category is the penalization-based method, which promotes sparsity by explicitly imposing regularization on the weights of the basis functions during optimization. For example, the Sparse Identification of Nonlinear Dynamics (SINDy) enforces sparsity by imposing $L1$-regularization on the weights of the basis functions, thus retaining only the most relevant basis functions~\cite{SINDY, Rosafalco2025, SINDywithControl, SINDY3}. Alternatively, Bayesian regression methods have gained increasing attention for their principled probabilistic framework, which enables uncertainty quantification while inherently supporting sparse selection. Particularly, Automatic Relevance Determination (ARD)~\cite{ARD, ARDS, ARD3} generalizes the concept by assigning individual Gaussian priors to each weight parameter. This formulation allows the model to automatically infer the relevance of each basis function through adaptive adjustment of its prior variance, leading to finer sparsity control. Beyond these two approaches, a variety of other methods have also been introduced into the field of sparse identification. The Perron-Frobenius operator methods \cite{PFO1, PFO2} and cluster-based reduced-order methods~\cite{order1, order2} enhance modeling sparsity and flexibility through data-driven operator representations while remaining sensitive to dimensionality and noise. Notably, reinforcement learning-based methods have also achieved progress in dynamical equation discovery~\cite{Li2025, bi-level, Holt2024}. For example, the Bi-level Identification of Equations (BILLIE)~\cite{bi-level} achieves state-of-the-art results in dynamical equation discovery. Nevertheless, its trial-and-error-based training approach incurs substantial computational costs.


Overall, although existing sparse identification methods have achieved notable progress, several limitations still remain:
\begin{enumerate}

  \item \textbf{Sequential data scenario:} Most sparse identification algorithms rely on previously collected historical datasets for batch training, which is computationally expensive and prevents efficient adaptation to sequentially arriving data. As a result, their application in online scenarios is limited.
  
  \item \textbf{Partially observable conditions:} In practical applications, systems are usually partially observable and noisy. Conventional methods typically assume access to accurate, full-state observations, which limits their direct applicability in practical applications.

\end{enumerate}

To address these limitations, this paper proposes a Sparse Kalman Identification (SKI) method that establishes a unified Bayesian sparsification framework based on the Augmented Kalman Filter (AKF). The SKI method seamlessly integrates Kalman recursive estimation and ARD-based sparsification into a single probabilistic framework, enabling dynamic model structure learning from sequential data. Specifically, to handle nonlinear and partially observable conditions, an augmented Unscented Kalman Filter (UKF)~\cite{UKF, UKF_CHOLESKY} is employed to ensure stable and accurate estimation of the system states and model parameters. Simultaneously, an online ARD-based sparsification mechanism is developed, which adaptively determines the relevance of basis functions by updating their prior variances in a Bayesian manner. The main contributions are summarized as follows:
\begin{enumerate}
    \item \textbf{Online ARD-based sparsification scheme integrated with AKF:} 
    A theoretically grounded online sparsification mechanism is developed based on the Augmented Kalman Filter (AKF), enabling relevance learning of basis functions directly from sequential data without relying on batch training. This design allows the model to adaptively achieve sparse structure selection in online scenario.
    
    \item \textbf{Kalman-Filter-like posterior update mechanism for model structure adaptation:} 
    To ensure that the learned relevance of basis functions can reflect the updated selection of the model structure, a Kalman-Filter-like posterior update mechanism is established, allowing the estimated states and model parameters to adapt dynamically to the updated ARD relevance.
    
    \item \textbf{Explicit gradient-descent formulation for efficient online updating:} 
    To enhance computational efficiency, an explicit gradient-descent formulation is derived for relevance updating, supporting online sparse identification with millisecond-level efficiency.
\end{enumerate}

Both simulation studies and real-world quadrotor experiments have been conducted to validate the effectiveness of the proposed method. The results demonstrate that, across three different sparsity selection tasks, the proposed Sparse Kalman Identification (SKI) method successfully identifies the relevant basis functions from a redundant candidate set, whereas the baseline algorithms fail to do so. Moreover, in the benchmark experiment, the proposed method outperforms the baseline approaches in terms of identification accuracy, achieving an $84.21\%$ improvement over the best baseline algorithm, while maintaining comparable computational efficiency (2.49~ms per step). These advantages make the method well-suited for practical online modeling and control scenarios.

The remainder of this paper is organized as follows. Section~\ref{sec:problem} provides a problem formulation, introducing the general formulation of nonlinear system identification under partial observation and reviewing representative methods. Section~\ref{sec:method} details the Sparse Kalman Identification (SKI) method, including the state-augmented UKF and the online ARD-based sparsification method. Section~\ref{sec:evaluation} presents experimental validation through simulation studies and real-world quadrotor experiments. Finally, Section~\ref{sec:conclusion} summarizes the main contributions of the research.

\section{Problem Formulation}
\label{sec:problem}

This section formulates the problem of dynamics identification for nonlinear and partially observable systems. Several representative data-driven methods and their typical implementations are briefly reviewed, followed by a discussion of the motivation for the proposed method.

\subsection{Formulation of the Dynamics Identification Problem}

Mathematically, a nonlinear dynamical system is typically described by the following discrete-time state-space model:

\begin{equation}
  \begin{aligned}
  \label{eq.discrete_state_space_model}
  \bm{x}_{t+1} &= \bm{F}(\bm{x}_t, \bm{u}_t, \bm{f}(\bm{x}_t, \bm{u}_t)) + \bm{w}_t\\
  \bm{y}_t &= \bm{h}(\bm{x}_t) + \bm{v}_t
  \end{aligned}
\end{equation}
where $\bm{x}_t \in \mathbb{R}^{d_x}$ and $\bm{u}_t \in \mathbb{R}^{d_u}$ denote the system state and input at time $t$, respectively. The function $\bm{F}: \mathbb{R}^{d_x} \times \mathbb{R}^{d_u} \times \mathbb{R}^{d_f} \to \mathbb{R}^{d_x}$ represents the state transition function, which may be nonlinear, and the function $\bm{f}: \mathbb{R}^{d_x} \to \mathbb{R}^{d_f}$ specifically indicates the unknown and potentially nonlinear term within the state transition, where $d_f$ is the dimension of the unknown term. In the observation equation, $\bm{h}: \mathbb{R}^{d_x} \to \mathbb{R}^{d_y}$ denotes the known observation function that may provide partial measurements (the full state $\bm{x}_t$ cannot be directly inferred from the observation $\bm{y}_t \in \mathbb{R}^{d_y}$). Finally, the process noise $\bm{w}_t \in \mathbb{R}^{d_x}$ and measurement noise $\bm{v}_t \in \mathbb{R}^{d_y}$ are assumed to be zero-mean Gaussian, i.e., $\bm{w}_t \sim \mathcal{N}(\bm{0}, \bm{Q})$ and $\bm{v}_t \sim \mathcal{N}(\bm{0}, \bm{R})$.

\subsection{Related Works and Issues}

For data-driven modeling approaches, it is typically assumed that system observation data over a time horizon are available for model identification, denoted as $\bm{y}_{1:t} = [\bm{y}_1, \bm{y}_2, \ldots, \bm{y}_t]$. In non-parametric data-driven methods, for example, the Deep Neural Network (DNN)-based methods, the unknown dynamics $\bm{f}(\cdot)$ are generally approximated by a parameterized neural network, i.e. $\bm{f}(\cdot) \approx \bm{f}_{\mathrm{DNN}}(\cdot)$. Alternatively, Gaussian Processes (GP) regression offers a non-parametric approach for modeling the unknown dynamics. In this framework, $\bm{f}(\cdot)$ is modeled as a stochastic process with a specified mean function $\bm{m}: \mathbb{R}^{d_x} \to \mathbb{R}^{d_f}$ and covariance function $\bm{K}: \mathbb{R}^{d_x} \times \mathbb{R}^{d_x} \to \mathbb{R}^{d_f \times d_f}$, i.e., $\bm{f}(\cdot) \sim \mathcal{GP}(\bm{m}(\cdot), \bm{K}(\cdot, \cdot))$. The covariance function $\bm{K}(\cdot, \cdot)$ is often chosen as a Radial Basis Function (RBF) kernel, capturing smoothness and correlation in the data. Model training involves computing the posterior distribution over functions conditioned on the observed data, enabling probabilistic predictions and uncertainty quantification~\cite{gp_1, ZHENG2025110049}. Despite the flexibility and expressive power, non-parametric data-driven methods typically require large volumes of training data and substantial computational resources for training. Consequently, they are often impractical for real-time applications.

On the other hand, parametric identification methods typically approximate the unknown function $\bm{f}(\cdot)$ as a linear combination of predefined basis functions. The weight parameters $\bm{\theta} \in \mathbb{R}^{d_\theta}$ of the basis functions are then identified by data-driven methods ($ d_\theta $ is the number of basis functions), resulting in the following representation:

\begin{equation}
  \begin{aligned}
  \bm{x}_{t+1} &= \bm{F}(\bm{x}_t, \bm{u}_t, \bm{\Phi}(\bm{x}_t, \bm{u}_t)^T\bm{\theta}) + \bm{w}_t\\
  \bm{y}_t &= \bm{h}(\bm{x}_t) + \bm{v}_t
  \end{aligned}
  \label{eq.parametric_state_space_model}
\end{equation}
where $\bm{\Phi}(\bm{x}_t, \bm{u}_t)=\left[{\phi}_i(\bm{x}_t, \bm{u}_t)\right]_{i = 1}^{d_\theta}$ is a column vector containing the evaluations of candidate basis functions ${\phi}_i: \mathbb{R}^{d_x} \times \mathbb{R}^{d_u} \to \mathbb{R}^{d_f}$ at the state $\bm{x}_t$ and input $\bm{u}_t$. In such parametric frameworks, dynamics identification is thus achieved by estimating $\bm{\theta}$. Methods such as Recursive Least Squares (RLS), Augmented Kalman Filter (AKF) and frequency-domain approaches can be directly employed to perform the data-driven estimation of $\bm{\theta}$~\cite{AKF,FTR1}. However, as the system dynamics become more complex, a large number of basis functions are required to ensure sufficient modeling accuracy. This expansion may lead to overfitting and increased sensitivity to noise when the data is insufficient. To address these issues, sparse identification methods have been developed to automatically select the most relevant basis functions~\cite{SINDY, BRR1, ARD, ARDS}. As a representative example, the Sparse Identification of Nonlinear Dynamics (SINDy) framework assumes that a system’s dynamics can be approximated as a sparse linear combination of candidate basis functions~\cite{SINDY, Rosafalco2025}.  
For a fully observable state vector $\bm{x}_k \in \mathbb{R}^{d_x}$ and control input $\bm{u}_k \in \mathbb{R}^{d_u}$ at sampled time $t=k$, 
the canonical formulation of the Sparse Identification of Nonlinear Dynamics (SINDy) method 
approximates the system dynamics as
\begin{equation}
  \dot{\bm{X}} \approx \bm{\Psi}(\bm{X}, \bm{U})\,\bm{\Xi}
  \label{eq:sindy_matrix_form}
\end{equation}
where $\dot{\bm{X}}\in\mathbb{R}^{N\times d_x}$ stacks the estimated state derivatives 
$\dot{\bm{x}}_k^T$ across all $N$ sampled data points. The matrix $\bm{\Psi}(\bm{X},\bm{U})\in\mathbb{R}^{N\times d_\Psi}$ denotes the library matrix, whose $k$-th row $\bm{\Psi}_{k,:} = \bm{\psi}(\bm{x}_k, \bm{u}_k)^T$ contains the evaluations of a predefined set of $d_\Psi$ candidate basis functions, 
$\bm{\psi}(\cdot):\mathbb{R}^{d_x}\times\mathbb{R}^{d_u}\rightarrow\mathbb{R}^{d_\Psi}$, 
at the $k$-th data sample. The coefficient matrix $\bm{\Xi}\in\mathbb{R}^{d_\Psi\times d_x}$ 
encodes the linear weights that combine the basis functions to reconstruct the system dynamics,
where the $i$-th column $\boldsymbol{\xi}^{(i)} = \bm{\Xi}_{:,i}$ corresponds to 
the coefficients governing the evolution of the $i$-th state component. The sparse identification objective is to determine the most parsimonious representation of the dynamics,
such that only the most relevant basis functions are retained in each column of $\bm{\Xi}$.
Given the sampled dataset 
$\{(\bm{x}_k, \bm{u}_k, \dot{\bm{x}}_k)\}_{k=1}^{N}$,
SINDy estimates the coefficients by solving an independent sparse regression problem for each state dimension $i=1,\ldots,d_x$:
\begin{equation}
  \min_{\boldsymbol{\xi}^{(i)}\in\mathbb{R}^{d_\Psi}} 
  \big\|
    \dot{\bm{X}}_{:,i} - \bm{\Psi}\boldsymbol{\xi}^{(i)}
  \big\|_2^2
  + \lambda_i \|\boldsymbol{\xi}^{(i)}\|_1
  \label{eq:sindy_lasso}
\end{equation}
where $\dot{\bm{X}}_{:,i}\in\mathbb{R}^{N}$ denotes the vector of time derivatives 
for the $i$-th state component,
and $\lambda_i>0$ is a regularization parameter controlling the sparsity level of the solution.
Eq.~\eqref{eq:sindy_lasso} is equivalent to a column-wise regression with $\ell_1$ regularization,
and alternative structured regularization schemes (e.g., group sparsity across columns)
can be employed to enforce shared support among multiple state dimensions. However, several critical limitations prevent direct application of classical SINDy to real-world scenarios. First, SINDy operates as a batch algorithm that depends on historical state data for each update. Therefore, it is not well suited for online learning, in which data arrive sequentially. Second, SINDy assumes full observation of the system state. In cases of partially observable systems, accurate construction of $\bm{\Psi}$ and estimation of $\dot{\bm{x}}_k$ become infeasible. 

To address these limitations, this paper proposes a new online sparse identification method, termed Sparse Kalman Identification (SKI). The method enables joint estimation of system states and weight parameters for nonlinear systems under noisy and partial observations, while incorporating ARD-based sparsification to prune irrelevant basis functions. The main contributions and algorithmic improvements are presented in the following section.

\section{Implementation of the SKI for Online Dynamics Identification}
\label{sec:method}


This section proposes the detailed implementation of the Sparse Kalman Identification (SKI) method to achieve real-time dynamics identification. The method includes the AKF-based online parameter estimation algorithm to handle partially observable conditions, the online ARD-based relevance learning mechanism to achieve sparsification through adapting prior variances, the Kalman-Filter (KF)-like posterior update mechanism to attain the updated model parameters and states estimation, and the explicit gradient-descent formulation for efficient computation.

\subsection{AKF-Based Online Parametric Identification}

To enable real-time parametric identification for partially observable systems, this work introduces the Augmented Kalman Filter (AKF) for parameter estimation. In this framework, the vector of weight parameters $\bm{\theta}$ is augmented into the state space to form an augmented state vector $\bm{\bar{x}}_t = [\bm{x}_t^T, \bm{\theta}^T]^T$. This formulation allows the AKF to estimate both the system state and the parameters simultaneously within a unified recursive filtering process. Since the AKF inherits the structure of the standard Kalman Filter, its performance depends on how the nonlinearities in the state transition and observation models are handled. To accommodate potentially strong nonlinear dynamics, nonlinear extensions of the AKF are adopted, including the Extended Kalman Filter (EKF) and the Unscented Kalman Filter (UKF). The discrete-time parametric identification model for the augmented state is thus formulated as follows:

\begin{equation}
  \begin{aligned}
  {\bm{\bar{x}}_{t+1}} &= \bm{\bar{F}}(\bm{\bar{x}}_t, \bm{u}_t)+     \begin{bmatrix}
    \bm{w}_t \\
    \bm{0}
  \end{bmatrix} \\
  \bm{\bar{F}}(\bm{\bar{x}}_t, \bm{u}_t) &=     
  \begin{bmatrix}
    \bm{F}(\bm{x}_t, \bm{u}_t, \bm{\Phi}(\bm{x}_t, \bm{u}_t)^T\bm{\theta}) \\
    \bm{\theta}
  \end{bmatrix}
  \end{aligned}
  \label{eq.tran_para}
\end{equation}
where the $\bm{\bar{F}}(\cdot)$ represents the discrete-time augmented state transition function, and $\bm{\Phi}(\bm{x}_t, \bm{u}_t)=\left[{\phi}_i(\bm{x}_t, \bm{u}_t)\right]_{i = 1}^{d_\theta}$ is a column vector of candidate basis functions ${\phi}_i: \mathbb{R}^{d_x} \to \mathbb{R}^{d_f}$. The initial distribution of $\bm{\bar{x}}_t$ is assumed to be Gaussian as follows:


\begin{equation}
  p(\bm{x}_0, \bm{\theta}) = \mathcal{N}\left(
    \begin{bmatrix}
      \bm{x}_0 \\
      \bm{\theta}
    \end{bmatrix}
    \Bigg|
    \begin{bmatrix}
      \bm{\mu}_0 \\
      \bm{m}_0
    \end{bmatrix},
    \begin{bmatrix}
      \bm{P}_0 & \bm{0} \\
      \bm{0} & \bm{S}_0
    \end{bmatrix}
  \right)
  \label{eq.init_para}
\end{equation}
where $\bm{\mu}_0$ and $\bm{m}_0$ denote the means, and $\bm{P}_0$ and $\bm{S}_0$ denote the covariance matrices of the initial system state $\bm{x}_0$ and the parameter vector $\bm{\theta}$, respectively. For brevity, we denote the mean vector and covariance matrix of the augmented state at time $t$ as $\bm{\xi}_t = [\bm{\mu}_t, \bm{m}_t]^T$ and $\bm{\Sigma}_t = \begin{bmatrix} \bm{P}_t & \bm{V}_t \\ \bm{V}_t^T & \bm{S}_t \end{bmatrix}$, where $\bm{V}_t$ captures the cross-covariance between $\bm{x}_t$ and $\bm{\theta}$. After formulating the augmented state transition function and the initial distribution, the EKF and UKF algorithm can be employed to estimate the augmented state:


\textbf{The augmented EKF algorithm} achieves nonlinear state estimation by locally linearizing nonlinear system state functions around the estimation, enabling the prediction and correction of the augmented state. Assuming that the mean and variance of the augmented state at time $t-1$ are known, the augmented state transition function $\bm{\bar{F}}(\cdot)$ can be approximated by a first-order Taylor expansion at the mean $\bm{\xi}_{t-1}^+$ as follows:
\begin{equation}
  \label{eq.ekf_approx}
  \bm{\bar{F}}(\bm{\bar{x}}_{t-1}, \bm{u}_{t-1}) \approx \bm{\bar{F}}(\bm{\xi}_{t-1}^+, \bm{u}_{t-1}) + \bm{J}_{\bm{\xi}, t-1} (\bm{\bar{x}}_{t-1} - \bm{\xi}_{t-1}^+)
\end{equation}

For simplicity, we denote $\bm{J}_{\bm{\xi}, t-1} = \left.\frac{\partial \bar{F}(\bm{\xi})}{\partial \bm{\xi}}\right|_{\bm{\xi}=\bm{\xi}_{t-1}^+}$ is the Jacobian matrix of $\bm{\bar{F}}(\cdot)$ with respect to $\bm{\xi}_{t-1}^+$. After getting the local linearization of the augmented state transition function, the prediction step of the augmented Kalman Filter can be expressed as:


\begin{equation}
  \label{eq.ekf_pred}
  \begin{aligned}
  \bm{\xi}^-_{t} &= \bm{\bar{F}}(\bm{\xi}_{t-1}^+, \bm{u}_{t-1}) \\
  \bm{\Sigma}^-_{t} &= \bm{J}_{\bm{\xi}, t-1} \bm{\Sigma}_{t-1} \bm{J}_{\bm{\xi}, t-1}^T + \bm{\bar{Q}}
  \end{aligned}
\end{equation}
where $\bm{\bar{Q}} = \begin{bmatrix} \bm{Q} & \bm{0} \\ \bm{0} & \bm{0} \end{bmatrix}$ is the augmented process noise covariance matrix. Upon receiving the current measurement $\bm{y}_{t}$ via the observation model, the correction step can be described as follows:


\begin{equation}
  \label{eq.ekf_correct}
  \begin{aligned}
  \bm{K}_t &= \bm{\Sigma}^-_t \bm{\bar{H}}_t^{T} \left( \bm{\bar{H}}_t \bm{\Sigma}^-_t \bm{\bar{H}}_t^{T} + \bm{R}_t \right)^{-1} \\
  \bm{\xi}^+_{t} &= \bm{\xi}^-_{t} + \bm{K}_{t} (\bm{y}_{t} - \bm{h}(\bm{x}_{t})) \\
  \bm{\Sigma}^+_{t} &= (\bm{I} - \bm{K}_{t} \bm{\bar{H}}_{t}) \bm{\Sigma}^-_{t}
  \end{aligned}
\end{equation}
where $\bm{K}_{t}$ is the Kalman gain, $\bm{\bar{H}}_{t} = [\left.\frac{\partial \bm{h}(\bm{\xi})}{\partial \bm{\xi}}\right|_{\bm{\xi}=\bm{\xi}_{t}^-}, \bm{0}]$ is the augmented Jacobian matrix of $\bm{h}(\cdot)$ with respect to $\bm{\xi}^-_{t}$, and $\bm{I}$ is the identity matrix. When the system dynamics exhibit high-order nonlinearity, the first-order linearization-based EKF may incur approximation errors, thereby hindering the accurate estimation of $\bm{\theta}$. To overcome this limitation, UKF is introduced.

\textbf{The augmented UKF algorithm} leverages the unscented transformation to more faithfully propagate the statistics of nonlinear systems. Specifically, instead of relying on Jacobian-based linearization, the UKF deterministically selects a set of sigma points that exactly capture the mean and covariance of a Gaussian distribution. By propagating these sigma points through the nonlinear functions, the UKF achieves an accuracy that is up to the third Taylor series order, while the EKF retains only first-order accuracy~\cite{UKF}. To further enhance numerical stability and computational efficiency, the UKF in this work is implemented using the Cholesky decomposition~\cite{UKF_CHOLESKY}. The pseudocode of the UKF algorithm is in Algorithm~\ref{alg:ukf1} (The detailed derivation is provided in Appendix~\ref{appendix:ukf_cholesky}):


\begin{algorithm}[htbp]
  \caption{Cholesky-form UKF}\label{alg:ukf1}
  \begin{algorithmic}[1] 
  \item[] \textbf{Input:} initial mean of the augmented state $\bm{\xi}_{0}=[\bm{\mu}_{0}, \bm{m}_{0}]^T$, initial covariance matrix $\bm{\Sigma}_{0} = \begin{bmatrix} \bm{P}_{0} & \bm{0} \\ \bm{0} & \bm{S}_{0} \end{bmatrix}$.
  
  \STATE Initialize the hyperparameter of UKF, including $\alpha$ (usually set to $1e\!-\!4 \le \alpha \le 1$), $\beta$ (=2 is optimal for Gaussian distribution), the dimension of sigma-points $L = d_x + d_\theta$. 
  
  \STATE Calculate the scale parameters $\lambda = L(\alpha^2-1)$, $\eta = \sqrt{L + \lambda}$ and the weight  
  $W_0^{(m)} = \lambda/(L+\lambda), \; W_0^{(c)}=\lambda/(L+\lambda) + (1-\alpha^2+\beta)$,  
  $W_i^{(m)}=W_i^{(c)}=1/(2L+2\lambda), \, i=1,2,\dots,2L$. 
  
  \STATE Sample sigma points for the augmented state:
      \begin{equation*}
        \bm{\chi}_{t-1} = \begin{bmatrix}
          \bm{\xi}_{t-1} 
          & \bm{\xi}_{t-1} + \eta \bm{U}_{t-1}
          & \bm{\xi}_{t-1} - \eta \bm{U}_{t-1}
      \end{bmatrix}
      \end{equation*}
      where $\bm{U}_{t-1}$ is the Cholesky factor of $\bm{\Sigma}_{t-1}$.
  
  \STATE Propagate the sigma points through the discretized augmented state transition function:
      \begin{equation*}
        \bm{\chi}_{t|t-1} = \bm{\bar{F}}(\bm{\chi}_{t-1}, \bm{u}_{t-1})
      \end{equation*}
      where matrix $\bm{\chi}_{t|t-1}$ collects the propagated sigma points as its columns.
  
  \STATE Calculate the Cholesky-form distribution of the predicted augmented state $\bm{\bar{x}}_{t}$:
      \begin{equation*}
      \begin{aligned}
          \bm \xi_{t}^- &= \sum_{i=0}^{2L} W_i^{(m)} \bm{\chi}_{i,t|t-1}, \\[4pt]
          \mathcal{Q}_x \mathcal{R}_x &= \left[ \sqrt{W_1^{(c)}}(\bm{\chi}_{1:2L,t|t-1} - \bm{\xi}_t^-),\; \bm{Q}^{1/2} \right],\qquad \bm{U}_t^- = \mathcal{R}_x^{T} \\[4pt]
          \bm{U}_t^- &= \mathrm{cholupdate}\left( \bm{U}_t^-,\; \bm{\chi}_{0,t|t-1} - \bm{\xi}_t^-,\; W_0^{(c)} \right)
      \end{aligned}
      \end{equation*}
      where $\bm{\chi}_{i,t|t-1}$ denotes the $i$-th column of $\bm{\chi}_{t|t-1}$, $\bm{Q}^{1/2}$ is the Cholesky factor of the process noise covariance matrix, and $\mathrm{cholupdate}(\cdot)$ is the Cholesky rank-one update algorithm~\cite{cholupdate}.
  \STATE Propagate the sigma points of $\bm{\bar{x}}_{t-1}$ through the observation function:
      \begin{equation*}
      \begin{aligned}
        \bm{\gamma}_{t} &= \bm{h}(\bm{\chi}_{t|t-1}),\qquad \bm{y}_t^- &= \sum_{i=0}^{2L} W_i^{(m)} \bm{\gamma}_{i,t}
      \end{aligned}
      \end{equation*}
      where matrix $\bm{\gamma}_{t}$ collects the measured sigma points as its columns.
  \STATE Calculate the Cholesky-form distribution of the predicted measurement :
      \begin{equation*}
      \begin{aligned}
        \mathcal{Q}_y \mathcal{R}_y &= \left[ \sqrt{W_1^{(c)}}(\bm{\gamma}_{1:2L,t} - \bm{y}_t^-),\; \bm{R}^{1/2} \right],\qquad \bm{U}_{y_t} = \mathcal{R}_y^{T}, \\[4pt]
        \bm{U}_{y_t} &= \mathrm{cholupdate}\left( \bm{U}_{y_t},\; \bm{\gamma}_{0,t} - \bm{y}_t^-,\; W_0^{(c)} \right)
      \end{aligned}
      \end{equation*}
      where $\bm{\gamma}_{i,t|t-1}$ denotes the $i$-th column of $\bm{\gamma}_{t|t-1}$, and $\bm{R}^{1/2}$ is the Cholesky factor of the measurement noise covariance matrix.
  \STATE Calculate the Kalman gain and correct distribution of the estimation:
      \begin{equation*}
      \begin{aligned}
        \bm{C}_t &= \sum_{i=0}^{2L} W_i^{(c)} (\bm{\chi}_{i,t|t-1} - \bm{\xi}_t^-)(\bm{\gamma}_{i,t} - \bm{y}_t^-)^{T},\qquad \bm{K}_t = \bm{C}_t (\bm{U}_{y_t} \bm{U}_{y_t}^{T})^{-1}, \\[4pt]
        \bm{\xi}_t^+ &= \bm{\xi}_t^- + \bm{K}_t (\bm{y}_t - \bm{y}_t^-),\qquad \bm{U}_t^+ = \mathrm{cholupdate}\left( \bm{U}_t^-,\; \bm{K}_t \bm{U}_{y_t},\; -1 \right)
      \end{aligned}
      \end{equation*}
      where $\bm{U}_t^+$ is the Cholesky factor of the corrected covariance matrix $\bm{\Sigma}^+_{t}$.
  \end{algorithmic}
\end{algorithm}

In summary, this subsection establishes a probabilistic framework (AKF) that unifies state estimation under partial observations with weight parameters updating. This integration lays the groundwork for online model learning. However, the AKF itself does not address how to determine an appropriate set of basis functions for representing system dynamics. The construction of the basis-function set $\bm{\Phi}$ is challenging due to limited prior knowledge. Too few basis functions may limit the model’s expressiveness, while too many may introduce redundancy and lead to overfitting. To overcome this limitation, it is desirable to design a sparse learning mechanism that can retain only the most relevant basis functions from sequential data. Such a mechanism would yield a compact yet accurate model, improving learning efficiency, interpretability, and generalization in online applications.


\subsection{Online ARD-Based Sparsification and Kalman Posterior Update Mechanism}


In this subsection, a new online ARD-based sparsification method is proposed. This method dynamically adjusts the prior variances of the weight parameters, enabling recursive sparsification without storing historical data.


As discussed in the introduction, to enhance the sparsity of parametric identification, a commonly used approach is $\ell_1$-regularization~\cite{LASSO}. However, these approaches are not well suited to online scenarios for two reasons. First, their performance depends heavily on the choice of penalty hyperparameters, which complicates real-time algorithm tuning and may impair generalization. Second, using an $\ell_1$ penalty destroys the conjugacy between the Gaussian prior and the updated posterior, resulting in a non-Gaussian posterior and making integrating with AKF more challenging. Considering these limitations of regularization-based methods, we adopt the Bayesian regression method, Automatic Relevance Determination (ARD), as the baseline algorithm for sparsity promotion. The ARD method assigns independent Gaussian priors to each weight parameter and updates their prior variances via maximum likelihood estimation. This preservation of Gaussian conjugacy provides a natural motivation to attempt integrating ARD into the AKF framework. To further illustrate this compatibility of ARD with AKF, the basic implementation structure of ARD is first revisited. In standard ARD, the Gaussian prior of each parameter is assigned as follows:
\begin{equation}
    p(\theta_i \mid s_i) = \mathcal{N}(\theta_i \mid 0, s_i)
\end{equation}
It is worth noting that the initial prior distributions of the weight parameters are assumed to be independent, which implies that the prior covariance matrix $\bm{S}_0 = \mathrm{diag}(\bm{s})$ is diagonal. These prior variances are then estimated by maximizing the likelihood function:


\begin{equation}
    \bm{s}^* = \arg\max_{\bm{s}} p(\bm{y}_{1:T} \mid \bm{s})
\end{equation}
which can be computed in closed form for linear-Gaussian models, or approximated using Monte Carlo or variational inference when closed-form evaluation is intractable~\cite{ARD,ARDS,ARD3}. In ARD, the hierarchical prior settings naturally induces sparsity: since the prior over the weight parameters is assumed to be zero-mean Gaussian, a larger prior variance implies a higher probability of the corresponding weight being nonzero, indicating stronger relevance of the associated basis function. In contrast, the variances of irrelevant basis functions shrink toward zero, effectively driving their weights toward zero, thus enabling an automatic, data-driven sparsification mechanism. However, since this likelihood-based algorithm requires access to the entire observation history $\bm{y}_{1:T}$ for updates, it cannot be directly applied to online scenarios in which data arrives sequentially. Therefore, to integrate the ARD mechanism into the AKF framework without incurring considerable computational overhead, a recursive ARD update scheme is needed. Nevertheless, this recursive reformulation introduces two main challenges. First, to improve memory efficiency, the entire observation history $\bm{y}_{1:T}$ is not stored, which results in the loss of historical information necessary for updates. Second, the posterior distribution depends on the prior, implying that any change in the prior should trigger a corresponding real-time posterior update, while such an updating strategy has not been well established.


To address this challenge, this paper proposesd an online ARD algorithm that operates under a unified probabilistic framework with the AKF. The method extracts historical information from the posterior distribution of the augmented state. This information is then used to perform maximum-likelihood updates of the prior variances. Moreover, by introducing a Kalman-Filter-like pseudo-observation correction mechanism, it establishes a recursive scheme that enables posterior updates consistent with the updated priors. For notational clarity, the posterior distribution before the ARD update is denoted as $q^\mathrm{old}(\bm{\bar{x}}_t)$, and the posterior distribution after the ARD update as $q^\mathrm{new}(\bm{\bar{x}}_t)$, specifically:


\begin{equation}
  \label{eq.q_prob}
  \begin{aligned}
  q^{\mathrm{old}}(\bm{\bar{x}}_t) &\approx p(\bm{x}_t,\, \boldsymbol{\theta} \mid \bm{y}_{1:t},\, \bm{S}_0^\mathrm{old})\\
  &= \mathcal{N}(\bm{\bar{x}_t} \mid \bm{\xi}_t^{\mathrm{old}}, \bm{\Sigma}_t^{\mathrm{old}})\\
  q^{\mathrm{new}}(\bm{\bar{x}}_t) &\approx p(\bm{x}_t,\, \boldsymbol{\theta} \mid \bm{y}_{1:t},\, \bm{S}_0^\mathrm{new})\\
  &= \mathcal{N}(\bm{\bar{x}_t} \mid \bm{\xi}_t^{\mathrm{new}}, \bm{\Sigma}_t^{\mathrm{new}})
  \end{aligned}
\end{equation}
where the posterior distribution of the augmented state $q(\bm{\bar{x}_t}) \approx p(\bm{x}_t, \bm{\theta} \mid \bm{y}_{1:t}, \bm{S}_0)$ can be decomposed according to the Bayesian theory as follows:


\begin{equation}
  \label{eq.bayes_eq}
  \begin{aligned}
    p(\bm{x}_t, \bm{\theta} \mid \bm{y}_{1:t}, \bm{S}_0^\mathrm{old}) &= \frac{p(\bm{\theta} \mid \bm{S}_0^\mathrm{old}) \, p(\bm{x}_t, \bm{y}_{1:t} \mid \bm{\theta})}{p(\bm{y}_{1:t} \mid \bm{S}_0^\mathrm{old})}\\
    p(\bm{x}_t, \bm{\theta} \mid \bm{y}_{1:t}, \bm{S}_0^\mathrm{new}) &= \frac{p(\bm{\theta} \mid \bm{S}_0^\mathrm{new}) \, p(\bm{x}_t, \bm{y}_{1:t} \mid \bm{\theta})}{p(\bm{y}_{1:t} \mid \bm{S}_0^\mathrm{new})}
  \end{aligned}
\end{equation}
where $p(\bm{\theta} \mid \bm{S}_0)$ denotes the Gaussian prior distribution of the parameter vector $\bm{\theta}$, which is specified as a zero-mean Gaussian distribution $\mathcal{N}(\bm{\theta} \mid \bm{0}, \bm{S}_0)$ following the ARD algorithm. It is important to note that the likelihood term $p(\bm{x}_t, \bm{y}_{1:t} \mid \bm{\theta})$ does not depend on the prior covariance $\bm{S}_0$. 

To update the prior covariance from $\bm{S}_0^\mathrm{old}$ to $\bm{S}_0^{\mathrm{new}}$, the likelihood term $p(\bm{x}_t, \bm{y}_{1:t} \mid \bm{\theta})$ that appears in both the first and second equations of Eq.~\eqref{eq.bayes_eq} is utilized to combine the two equations, leading to a relationship between $\bm{S}_0^\mathrm{old}$ and $\bm{S}_0^{\mathrm{new}}$. Specifically, by reformulating the first equation of Eq.~\eqref{eq.bayes_eq}, the likelihood term can be expressed as:


\begin{equation}
  \label{eq.likelihood_exp}
  \begin{aligned}
    p(\bm{x}_t, \bm{y}_{1:t} \mid \bm{\theta}) &= \frac{p(\bm{x}_t, \bm{\theta} \mid \bm{y}_{1:t}, \bm{S}_0^\mathrm{old})  p(\bm{y}_{1:t} \mid \bm{S}_0^\mathrm{old})}{p(\bm{\theta} \mid \bm{S}_0^\mathrm{old})} \\ 
    &\propto \frac{q^{\mathrm{old}}(\bm{\bar{x}}_t)  p(\bm{y}_{1:t} \mid \bm{S}_0^\mathrm{old})}{p(\bm{\theta} \mid \bm{S}_0^\mathrm{old})}
  \end{aligned}
\end{equation}
By substituting Eq.~\eqref{eq.likelihood_exp} into the second equation of Eq.~\eqref{eq.bayes_eq}, we obtain a recursive update formula for the posterior distribution:


\begin{equation}
  \label{eq.recursive_update1}
  \begin{aligned}
  q^{\mathrm{new}}(\bm{\bar{x}}_t)
  &=
  \frac{
    p(\bm{y}_{1:t} \mid \bm{S}_0^\mathrm{old})
  }{
    p(\bm{y}_{1:t} \mid \bm{S}_0^\mathrm{new})
  }\frac{
    p(\bm{\theta} \mid \bm{S}_0^\mathrm{new})
  }{
    p(\bm{\theta} \mid \bm{S}_0^\mathrm{old})
  }q^{\mathrm{old}}(\bm{\bar{x}}_t) \\
  &\propto 
  \frac{
    \mathcal{N}(\bm{\theta} \mid \bm{0}, \bm{S}_0^\mathrm{new})
  }{
    \mathcal{N}(\bm{\theta} \mid \bm{0}, \bm{S}_0^\mathrm{old})
  }q^{\mathrm{old}}(\bm{\bar{x}}_t)
  \end{aligned}
\end{equation}
where the ratio of two Gaussian distributions can be rewritten as a new Gaussian distribution:
\begin{equation}
  \label{eq.gaussian_composition}
  \frac{
    \mathcal{N}(\bm{\theta} \mid \bm{0}, \bm{S}_0^\mathrm{new})
  }{
    \mathcal{N}(\bm{\theta} \mid \bm{0}, \bm{S}_0^\mathrm{old})
  }
  =
  C\mathcal{N}(\bm{\theta} \mid \bm{0}, \Delta \bm{S}_0)
\end{equation}
where the constant $C = (2\pi)^{n_\theta/2} \frac{|\bm{S}_0^\mathrm{old}|^{1/2} |\Delta \bm{S}_0|^{1/2}}{|\bm{S}_0^\mathrm{new}|^{1/2}}$, the covariance matrix $\Delta \bm{S}_0 = \left[ (\bm{S}_0^{\mathrm{new}})^{-1} - (\bm{S}_0^{\mathrm{old}})^{-1} \right]^{-1}$.  This formulation demonstrates that, after updating the prior covariance from $\bm{S}_0^{\mathrm{old}}$ to $\bm{S}_0^{\mathrm{new}}$, the distribution of the augmented state can be recursively updated without requiring access to historical data:
\begin{equation}
  \label{eq.recursive_update2}
  q^{\mathrm{new}}(\bm{\bar{x}}_t)
  \propto \mathcal{N}(\bm{\theta} \mid \bm{0}, \Delta \bm{S}_0)q^{\mathrm{old}}(\bm{\bar{x}}_t)
\end{equation}

It can be observed that the form of Eq.~\eqref{eq.recursive_update2} closely resembles the correction step in the Kalman filter framework. Motivated by this analogy, we aim to leverage the Kalman filter theory to explicitly derive the posterior update equation. The core idea is to treat $q^{\mathrm{old}}(\bm{\bar{x}}_t)$ as the Kalman-prior distribution before correction, and interpret the term $\mathcal{N}(\bm{\theta} \mid \bm{0}, \Delta \bm{S}_0)$ as a pseudo-observation model for the parameter vector $\bm{\theta}$. Specifically, we utilize the well-known symmetry between the random variable and the mean in the Gaussian distribution, i.e., $\mathcal{N}(\bm{\theta} \mid \bm{0}, \Delta \bm{S}_0) = \mathcal{N}(\bm{0} \mid \bm{\theta}, \Delta \bm{S}_0)$. This allows us to reinterpret $\mathcal{N}(\bm{\theta} \mid \bm{0}, \Delta \bm{S}_0)$ as corresponding to a pseudo-observation on $\bm{\theta}$, which is equivalent to $p(\tilde{\bm{y}}_t = \bm{0} \mid \bar{\bm{x}}_t) = \mathcal{N}(\boldsymbol{\theta} \mid \bm{0}, \Delta \bm{S}_0)$. Further, this pseudo-observation can be equivalently formulated as an observation model of the augmented state $\bm{\bar{x}}_t$:


\begin{equation}
  \begin{aligned}
  &\tilde{\bm{y}}_t = \bm{\bar{H}}_p \bar{\bm{x}}_t + \tilde{\bm{v}} \qquad \tilde{\bm{v}} \sim \mathcal{N}(\bm{0}, \Delta \bm{S}_0),\\
  &\bm{\bar{H}}_p = \left[\begin{array}{cc} \bm{0} & \bm{I}_{d_\theta} \end{array}\right]
  \end{aligned}
  \label{eq.pesudo}
\end{equation}
where $\tilde{\bm{y}}_t$ is a pseudo measurement. Consequently, Eq.~\eqref{eq.recursive_update2} can be reformulated as a Kalman correction equation under the pseudo-observation formulation:
\begin{equation}
  \label{eq.recursive_pesudo}
  \begin{aligned}
  q^{\mathrm{new}}(\bm{\bar{x}}_t)
  &\propto 
  p(\tilde{\bm{y}}_t = \bm{0} \mid \bar{\bm{x}}_t)q^{\mathrm{old}}(\bm{\bar{x}}_t)
  \end{aligned}
\end{equation}
which enables the posterior update step to follow the Kalman observation correction structure:


\begin{equation}
  \label{eq.kf-like_update}
  \begin{aligned}
    &\bm{G} = \bm{\Sigma}_{t}^{\mathrm{old}} \bm{\bar{H}}_p^{T} \left( \bm{S}_{t}^{\mathrm{old}} + \Delta\bm{S_0} \right)^{-1} \\
    &\bm{\xi}_{t}^{\mathrm{new}} = \bm{\xi}_{t}^{\mathrm{old}} - \bm{G} \bm{\xi}_{t}^{\mathrm{old}} \\
    &\bm{\Sigma}_{t}^{\mathrm{new}} = \bm{\Sigma}_{t}^{\mathrm{old}} - \bm{G} \bm{\bar{H}}_p \bm{\Sigma}_{t}^{\mathrm{old}}
  \end{aligned}
\end{equation}

In summary, this subsection formulates a Bayesian principle for updating posterior when the prior is updated, and derives a KF-like update equation to attain the mean and covariance of the new posterior distribution. However, the marginal likelihood function $p(\bm{y}_{1:t} \mid \bm{S}_0^\mathrm{new})$ is not explicitly given here, which should be computed to serve as the basis for updating the prior variance. Therefore, the next step is to provide an explicit formula for computing the marginal likelihood (or loss) function and to derive an optimization algorithm for updating the prior variance.


\subsection{Online ARD Relevance Update with Explicit Gradient Formulas}


To enable efficient updating of prior variance via maximum marginal likelihood, this subsection derives the marginal likelihood function and introduces a gradient-based updating approach, alongside explicit gradient descent update equations for real-time and efficient prior variance adaptation.

To clearly extract the marginal likelihood function, we first need to manipulate Eq.~\eqref{eq.recursive_update1}. By substituting the merged Gaussian result from Eq.~\eqref{eq.recursive_update2} into the first equation of Eq.~\eqref{eq.recursive_update1}, we obtain:
\begin{equation}
  \label{eq.likelihood_ratio}
  \begin{aligned}
    q^{\mathrm{new}}(\bm{\bar{x}}_t)
    &=
    C\frac{
      p(\bm{y}_{1:t} \mid \bm{S}_0^\mathrm{old})
    }{
      p(\bm{y}_{1:t} \mid \bm{S}_0^\mathrm{new})
    }\mathcal{N}(\bm{\theta} \mid \bm{0}, \Delta \bm{S}_0)q^{\mathrm{old}}(\bm{\bar{x}}_t)
  \end{aligned}
\end{equation}
Integrating both sides of Eq.~\eqref{eq.likelihood_ratio} to obtain the marginal distribution, we obtain:

\begin{equation}
  \label{eq.likelihood_ratio_marginal}
  \begin{aligned}
    1
    =
    C\frac{
      p(\bm{y}_{1:t} \mid \bm{S}_0^\mathrm{old})
    }{
      p(\bm{y}_{1:t} \mid \bm{S}_0^\mathrm{new})
    } \int \mathcal{N}(\bm{\theta} \mid \bm{0}, \Delta \bm{S}_0)q^{\mathrm{old}}(\bm{\bar{x}}_t)\mathrm{d}\bm{\bar{x}}_t \\
    \frac{
      p(\bm{y}_{1:t} \mid \bm{S}_0^\mathrm{new})
    }{
      p(\bm{y}_{1:t} \mid \bm{S}_0^\mathrm{old})
    } = C\int \mathcal{N}(\bm{\theta} \mid \bm{0}, \Delta \bm{S}_0)q^{\mathrm{old}}(\bm{\bar{x}}_t)\mathrm{d}\bm{\bar{x}}_t
  \end{aligned}
\end{equation}

The numerator on the left of Eq.~\eqref{eq.likelihood_ratio_marginal} corresponds to the marginal likelihood of $\bm{S}_0^{\mathrm{new}}$ given past observations, which indicates how likely the observed data are under the updated prior covariance. Meanwhile, the right-hand side does not explicitly depend on the historical observations; instead, the information from past data has already been incorporated into the posterior distribution $q^{\mathrm{old}}(\bm{\bar{x}}_t)$. For notational convenience, we denote $\ell(\bm{S}_0^\mathrm{new}) = \log\frac{p(\bm{y}_{1:t} \mid \bm{S}_0^\mathrm{new})}{p(\bm{y}_{1:t} \mid \bm{S}_0^\mathrm{old})}$. Furthermore, to evaluate the required integral, we substitute the pseudo-observation model $p(\tilde{\bm{y}}_t = \bm{0} \mid \bm{\bar{x}}_t) = \mathcal{N}(\boldsymbol{\theta} \mid \bm{0}, \Delta \bm{S}_0)$ in Eq.~\eqref{eq.pesudo}, thus constructing the marginal integral over the joint distribution as follows:

\begin{equation}
  \label{eq.ard_loss_pri}
  \begin{aligned}
    \ell(\bm{S}_0^\mathrm{new}) &= \log C \int p(\tilde{\bm{y}}_t=\bm{0} | \bar{\bm{x}}_t)q^{\mathrm{old}}(\bm{\bar{x}}_t)\mathrm{d}\bm{\bar{x}}_t\\
    &= \log C \int p(\tilde{\bm{y}}_t=\bm{0}, \bar{\bm{x}}_t)\mathrm{d}\bm{\bar{x}}_t\\
    &= \log C + \log p(\tilde{\bm{y}}_t = \bm{0}) \\
  \end{aligned}
\end{equation}
since $\bm{\bar{x}}_t$ follows a Gaussian distribution:
$
\mathcal{N}(\bm{\bar{x}}_t\mid\boldsymbol{\xi}_t^\mathrm{old}, \boldsymbol{\Sigma}_t^\mathrm{old}) = \mathcal{N}\left(
  \begin{bmatrix}
    \bm{x}_t \\
    \bm{\theta}_t
  \end{bmatrix}
  \Bigg|
  \begin{bmatrix}
    \bm{\mu}_t^\mathrm{old} \\
    \bm{m}_t^\mathrm{old}
  \end{bmatrix},
  \begin{bmatrix}
    \bm{P}_t^\mathrm{old} & \bm{V}_t^\mathrm{old} \\
    (\bm{V}_t^{\mathrm{old}})^T & \bm{S}_t^\mathrm{old}
  \end{bmatrix}
\right)
$, according to the Gaussian linear transformation theorem, $\tilde{\bm{y}}$ also follows a Gaussian distribution $\tilde{\bm{y}}_t \sim \mathcal{N}(\bm{m}_{t}^\mathrm{old},\Delta\bm{S}_0+\bm{S}_{t}^\mathrm{old})$, thus:
\begin{equation}
  \label{eq.ard_loss_pri2}
    \ell(\bm{S}_0^\mathrm{new}) = \log C + \log\mathcal{N}(\bm{0}|\bm{m}_{t}^\mathrm{old},\Delta\bm{S}_0+\bm{S}_{t}^\mathrm{old})
\end{equation}

Based on the above results, by omitting terms in the expression that are independent of $\bm{S}_0^{\mathrm{new}}$, the practical ARD optimization objective is to minimize the following loss function (for a detailed derivation of this loss function, please refer to Appendix~\ref{appendix:ard_loss}):
\begin{equation}
  \label{eq.ard_loss_final}
  \begin{aligned}
    &\mathcal{L} = \mathcal{L}_{1} + \mathcal{L}_{2} \\
    &\mathcal{L}_{1} = (\bm{m}_{t}^{\mathrm{old}})^{T} \left(\bm{S}_{t}^{\mathrm{old}} + \Delta\bm{S}_0\right)^{-1} \bm{m}_{t}^{\mathrm{old}} \\
    &\mathcal{L}_{2} = \log \left| \bm{S}_0^{\mathrm{new}} + \left[\bm{I}_{d_\theta} - \bm{S}_0^{\mathrm{new}} (\bm{S}_0^{\mathrm{old}})^{-1}\right] \bm{S}_{t}^{\mathrm{old}} \right|
  \end{aligned}
\end{equation}

Furthermore, to enable efficient adaptation of the prior variances, this work further derives an analytical gradient-based updating equation for adapting the prior variances. To ensure the positivity of the variances, each prior variance $s_i$ is parameterized via the softplus transformation: $s_i = \log(1 + \exp(\tilde{s}_i))$, where $[\tilde{s}_1, \ldots, \tilde{s}_{d_\theta}]^T$ is the vector of unconstrained hyperparameters (the prior variances). The gradient of the loss function $\mathcal{L}$ with respect to $\tilde{s}_i$ is computed as follows:
\begin{equation}
  \sigma(x):=\frac{1}{1+e^{-x}},\qquad
  \frac{\partial s_i}{\partial \tilde{s}_i}=\sigma(\tilde{s}_i)
\end{equation}
 where $\sigma(\cdot)$ denotes the sigmoid function. Let $\bm{S}_0^{\mathrm{new}} = \operatorname{diag}(\bm{s})$, $\bm{A} = (\bm{S}_{t}^{\mathrm{old}}+\Delta\bm{S}_0)^{-1}$, and $\bm{M} = \bm{S}_0^{\mathrm{new}} + [\bm{I}_{d_\theta}-\bm{S}_0^{\mathrm{new}}(\bm{S}_0^{\mathrm{old}})^{-1}]\bm{S}_{t}^{\mathrm{old}}$. The gradient of $\mathcal{L}$ with respect to $\tilde{s}_i$ consists of two terms:
\begin{equation}
  \begin{aligned}
  \frac{\partial \mathcal{L}}{\partial \tilde{s}_i}
  &= \frac{\partial \mathcal{L}_1}{\partial \tilde{s}_i}
  + \frac{\partial \mathcal{L}_2}{\partial \tilde{s}_i} \\
  \frac{\partial \mathcal{L}_1}{\partial s_i}
  &= -[q_i]^2 \\
  \frac{\partial \mathcal{L}_2}{\partial s_i}
  &= [\bm{M}^{-1}]_{ii} \left(1 - \frac{[\bm{S}_t^{\mathrm{old}}]_{ii}}{[\bm{S}_0^{\mathrm{old}}]_{ii}}\right)
  \end{aligned}
\end{equation}
where $q_i$ is the $i$-th component of the vector $\bm{q} = \bm{A}\bm{m}_{t}^{\mathrm{old}}$, $[\cdot]_{ii}$ denotes the $i$-th diagonal element of the matrix. Applying the chain rule, the total gradient is:
\begin{equation}
  \frac{\partial \mathcal{L}}{\partial \tilde{s}_i} =
    \sigma(\tilde{s}_i)
    \left(
      [\bm{M}^{-1}]_{ii} \left(1 - \frac{[\bm{S}_t^{\mathrm{old}}]_{ii}}{[\bm{S}_0^{\mathrm{old}}]_{ii}}\right)
      - [q_i]^2
    \right)
\end{equation}

At each optimization step, the ARD hyperparameter vector $\bm{s}$ is updated using the gradient-based method:
\begin{equation}
  \bm{s} \leftarrow \bm{s} - \eta_{\mathrm{hp}}\, \nabla_{\bm{s}}\mathcal{L}(\bm{s})
\end{equation}
where $\eta_{\mathrm{hp}}$ denotes the learning rate. This gradient-based Automatic Relevance Determination (ARD) update enables online, data-driven refinement of the relevance prior within the AKF framework, thereby promoting sparsity and improving generalization performance.

\subsection{Framework of the Proposed Method}

\begin{figure*}[htbp]
  \centering
  \includegraphics[width=0.92\textwidth]{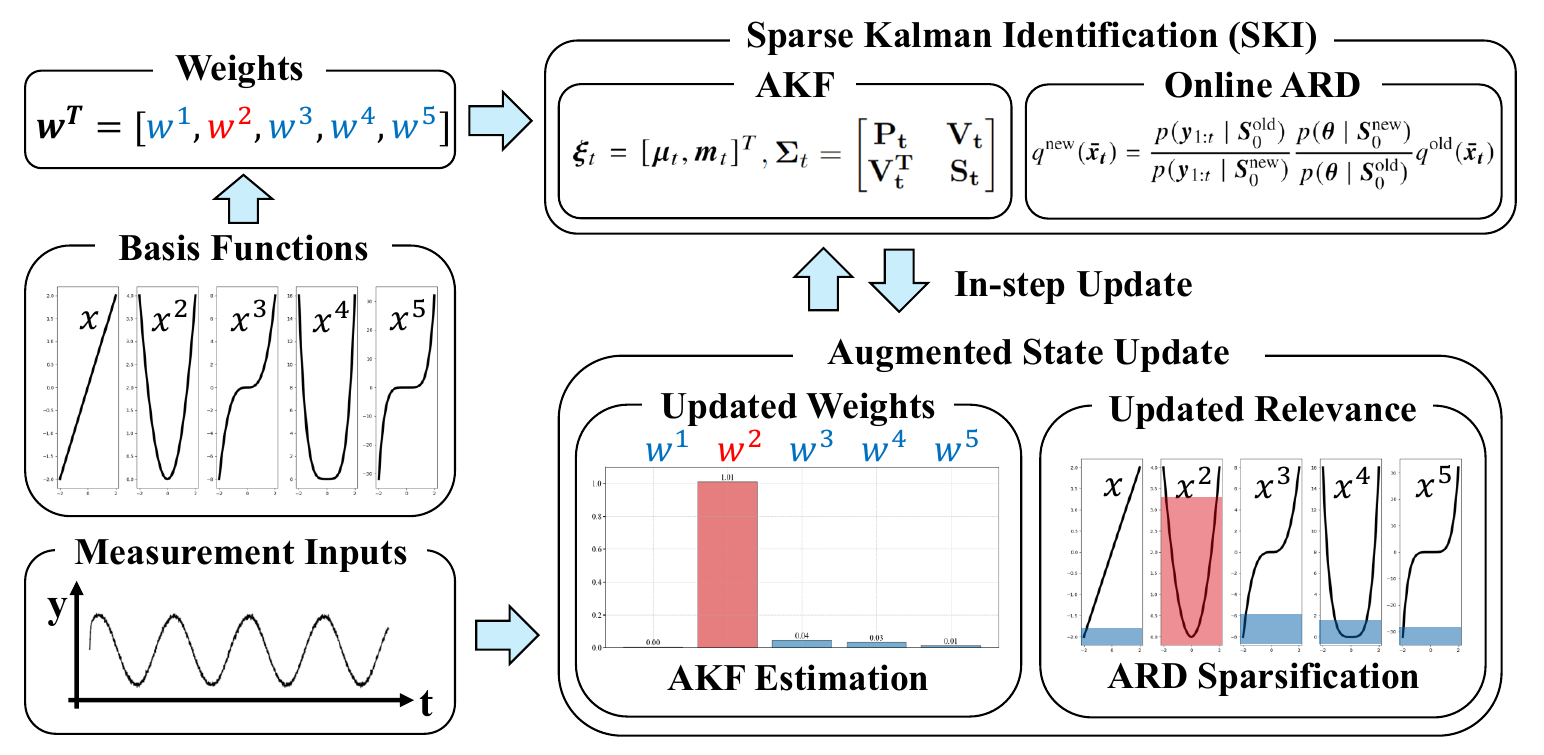}
  \caption{Overview of the proposed algorithm: The weight parameters of basis functions are augmented into the state, enabling joint online estimation and adaptive sparsity via ARD within the AKF framework.}
  \label{fig:framework}
\end{figure*}

In summary, the proposed Sparse Kalman Identification (SKI) method realizes sparse, interpretable, and efficient dynamics identification. First, by augmenting the state space to include both the physical states and the weight parameters, the algorithm enables joint estimation through the EKF or UKF, thus providing robust performance even under noisy and partially observable conditions. Second, the SKI method incorporates an ARD prior update mechanism, which adaptively adjusts the relevance of basis functions in real time. This is accomplished through a gradient-based update driven directly by the incoming current measurement, thereby eliminating the need to retain historical data. Third, after each ARD prior adjustment, a KF-like posterior update step is performed to update the model parameters (or model structure) and states estimation, achieving seamless integration of the AKF and ARD components while maintaining both real-time feasibility and sparsity. 

The algorithmic procedure of the proposed method is described in the form of pseudocode~\ref{alg:frame} and illustrated in Fig.~\ref{fig:framework}. The pseudocode systematically outlines the initialization, online AKF-based parameters estimation, and online ARD sparsification steps.

\begin{algorithm}[h]
  \caption{Core Procedure of the Proposed Method}\label{alg:frame}
  \begin{algorithmic}[1]
    \item[] \textbf{Input:} Set the initial state vector $\bm{x}_0$, state covariance matrix $\bm{P}_0$, parameter vector $\bm{\theta}_0$, parameter covariance matrix $\bm{S}_0$ (diagonal, initialized from ARD hyperparameters $\bm{s}_0$), the state-parameter covariance matrix $\bm{V}_0$, process noise covariance $\bm{Q}$, and measurement noise covariance $\bm{R}$.
    \REPEAT
        \STATE Acquire measurement $\bm{y}_t$.
        \STATE \textbf{UKF Update:} Propagate and correct the mean and covariance of $\bm{x}$ and $\bm{\theta}$ using Algorithm~\ref{alg:ukf1} with measurement $\bm{y}_t$.
        \FOR{$k = 1$ \textbf{to} $N_{\mathrm{hp}}$}
            \STATE Compute ARD loss $\mathcal{L}_{\mathrm{ARD}}$ based on current $\bm{\theta}_{t|t}$ and $\bm{S}_0$ using ~\eqref{eq.ard_loss_final}.
            \STATE Update ARD hyperparameters $\bm{s}_0$ via gradient-based optimization.
            \STATE Update parameter prior covariance $\bm{S}_0$ according to new $\bm{s}_0$.
        \ENDFOR
        \STATE Update the posterior to match the new prior using ~\eqref{eq.kf-like_update}
    \UNTIL{operation ends}
  \end{algorithmic}
\end{algorithm}

The computational complexity of the proposed algorithm is primarily determined by two components: the augmented UKF and the Online-ARD. In the Cholesky-form UKF, the most computationally intensive operation is the Cholesky decomposition of the augmented covariance matrix, whose dimension is $n_z = d_x + d_\theta$ (where $d_x$ and $d_\theta$ denote the state and parameter dimensions, respectively). This operation incurs a computational complexity of $\mathcal{O}(n_z^3)$, which typically dominates the overall computational cost of the UKF.

With the introduction of the Online-ARD mechanism, additional computations are required for the online optimization of hyperparameters governing the prior variances of the parameters $\bm{\theta}$. The principal computational burden in this module also arises from matrix operations, within the parameter subspace, with complexity on the order of $\mathcal{O}(n_\theta^3)$ per optimization step. If $N_{\mathrm{hp}}$ denotes the number of hyperparameter optimization steps performed at each filtering iteration, the total additional complexity introduced by Online-ARD is approximately $N_{\mathrm{hp}} \cdot \mathcal{O}(n_\theta^3)$.

In summary, although the incorporation of Online-ARD inevitably increases the overall computational cost per iteration, this increase remains acceptable in practice as long as the parameter dimension $n_\theta$ is not excessively large. Therefore, the proposed method maintains a reasonable balance between sparsification capability and computational efficiency, and the computational overhead introduced by Online-ARD is manageable for problems with moderate parameter dimensionality. Finally, the theoretical advantages of the Sparse Kalman Identification (SKI) algorithm will be rigorously validated in the following section, through comprehensive numerical simulations and real-world experiments.

\section{Evaluation}
\label{sec:evaluation}

This section provides experimental validation of the Sparse Kalman Identification (SKI) method under three representative scenarios. Specifically, Subsection~\ref{subsec:exp1} assesses the method on the numerical WingRock synthetic dynamics benchmark for system identification, validating the accuracy and efficiency of the proposed method. Subsection~\ref{subsec:exp2} evaluates the method's performance in a numerical quadrotor simulation based on the Gazebo platform, validating the sparsification adaptability of the proposed method under complex flight dynamics. Subsection~\ref{subsec:exp3} further demonstrates the sparsification adaptability using a physical quadrotor UAV experiment, thereby verifying its practical effectiveness in real-world applications. 

All simulation experiments are performed on a desktop workstation equipped with an Intel(R) Core(TM) i7-14700K 3.40 GHz CPU and 32 GB of RAM, using Python 3.9. Details of the hardware configuration of the physical quadrotor platform are provided in Section~\ref{subsec:exp3}.

\subsection{Numerical Simulation Experiment on the WingRock System}
\label{subsec:exp1}

The first simulation experiment utilizes the WingRock dynamical model, a typical benchmark for evaluating nonlinear and partially observable system identification~\cite{wingrock1, wingrock2GP, wingrock3}. The discrete-time dynamics of the WingRock system are described by the following equations:
\begin{equation}
    \begin{aligned}
        \dot{\theta}_t &= p_t \\
        \dot{p}_t &= L \Delta d + \Delta(\theta_t, p_t)
    \end{aligned}
    \label{eq:wingrock}
\end{equation}
where $\theta_t$ denotes the roll angle (in degrees), $p_t$ represents the roll rate (in degrees per second), $\Delta d$ is the control input (aileron deflection in degrees), and $L$ is the input gain, whose true value is set as $L = 3~\si{\per\second\squared}$. The term $\Delta(\theta_t, p_t)$ encapsulates the system's nonlinear uncertainty and is modeled as a weighted sum of nonlinear basis functions:
\begin{equation}
    \Delta(\theta_t, p_t) = \bm{\Phi}(\theta_t, p_t) \bm{w}
    \label{eq:uncertainty}
\end{equation}
where the basis function vector $\bm{\Phi}(\theta_t, p_t)$ comprises a collection of nonlinear terms, namely: a constant bias, the roll angle $\theta_t$, the roll rate $p_t$, the interaction term $|\theta_t|p_t$, the quadratic term $|p_t|p_t$, and the cubic term $\theta_t^3$:
\begin{equation}
    \bm{\Phi}(\theta_t, p_t) = \left[ 1, \theta_t, p_t, |\theta_t|p_t, |p_t|p_t, \theta_t^3 \right]
    \label{eq:phi}
\end{equation}
and $\bm{w} \in \mathbb{R}^{6 \times 1}$ is the weight vector to be identified, whose true values are given by $w_0 = 0.8\,\si{deg \cdot s^{-2}}$, $w_1 = 0.2314\,\si{s^{-2}}$, $w_2 = 0.6918\,\si{s^{-1}}$, $w_3 = -0.6245\,\si{deg^{-1}s^{-1}}$, $w_4 = 0.0095\,\si{deg^{-1}}$, and $w_5 = 0.0214\,\si{deg^{-2}\,s^{-2}}$ ~\cite{wingrock2GP}. The system is partially observable: only the roll angle $\theta_t$ is measured, and the measurement is corrupted by additive Gaussian white noise with a standard deviation of $0.1$ degrees.

\begin{figure*}[h]
  \centering
  \includegraphics[width=0.86\textwidth]{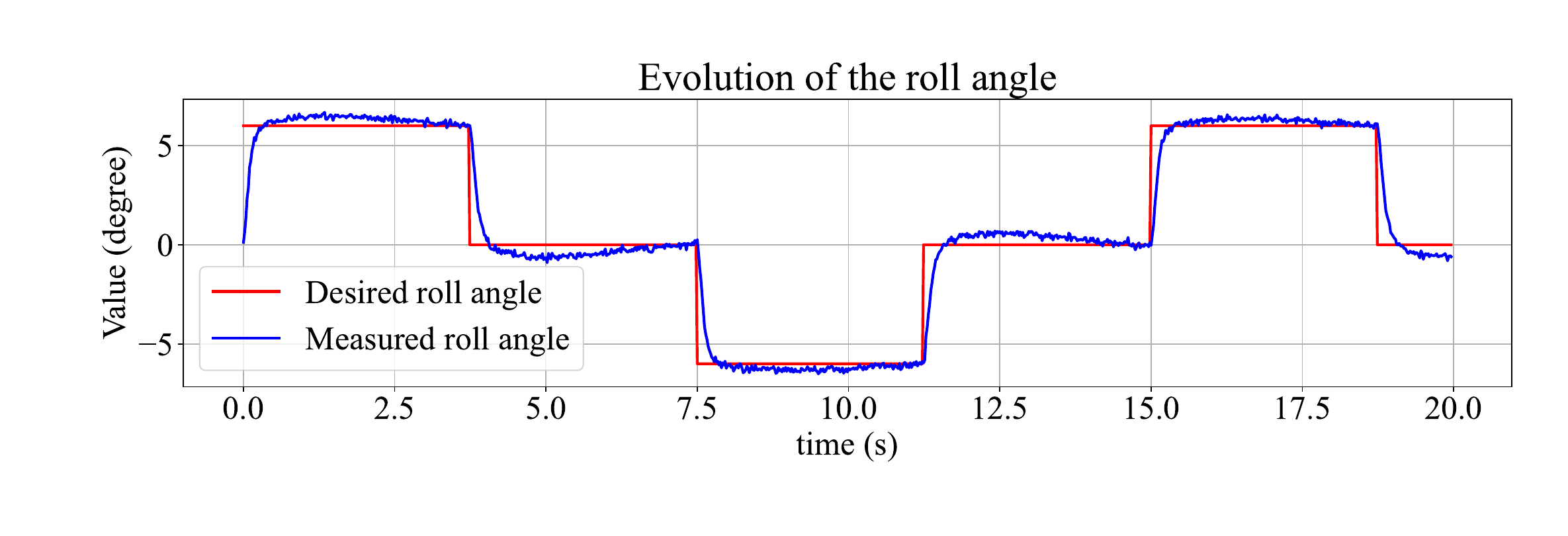}
  \caption{Evolution of the roll angle in the WingRock simulation experiment.}
  \label{fig:exp1_3}
\end{figure*}

The simulation is performed at a sampling frequency of 50~Hz for a total duration of 15 seconds. The control objective is to track a time-varying reference roll angle using a PID controller. The reference trajectory consists of a sequence of stepwise square waves, as illustrated in Fig.~\ref{fig:exp1_3}. To rigorously assess the parameter identification performance, we first conduct a comparative study among four algorithms: SINDy, the augmented Extended Kalman Filter, the augmented Unscented Kalman Filter, and our Sparse Kalman Identification (SKI) algorithm. It should be noted that, for SINDy, the angular velocity and angular acceleration are obtained via numerical differentiation.

\begin{figure*}[h]
  \centering
  \includegraphics[width=1.0\textwidth]{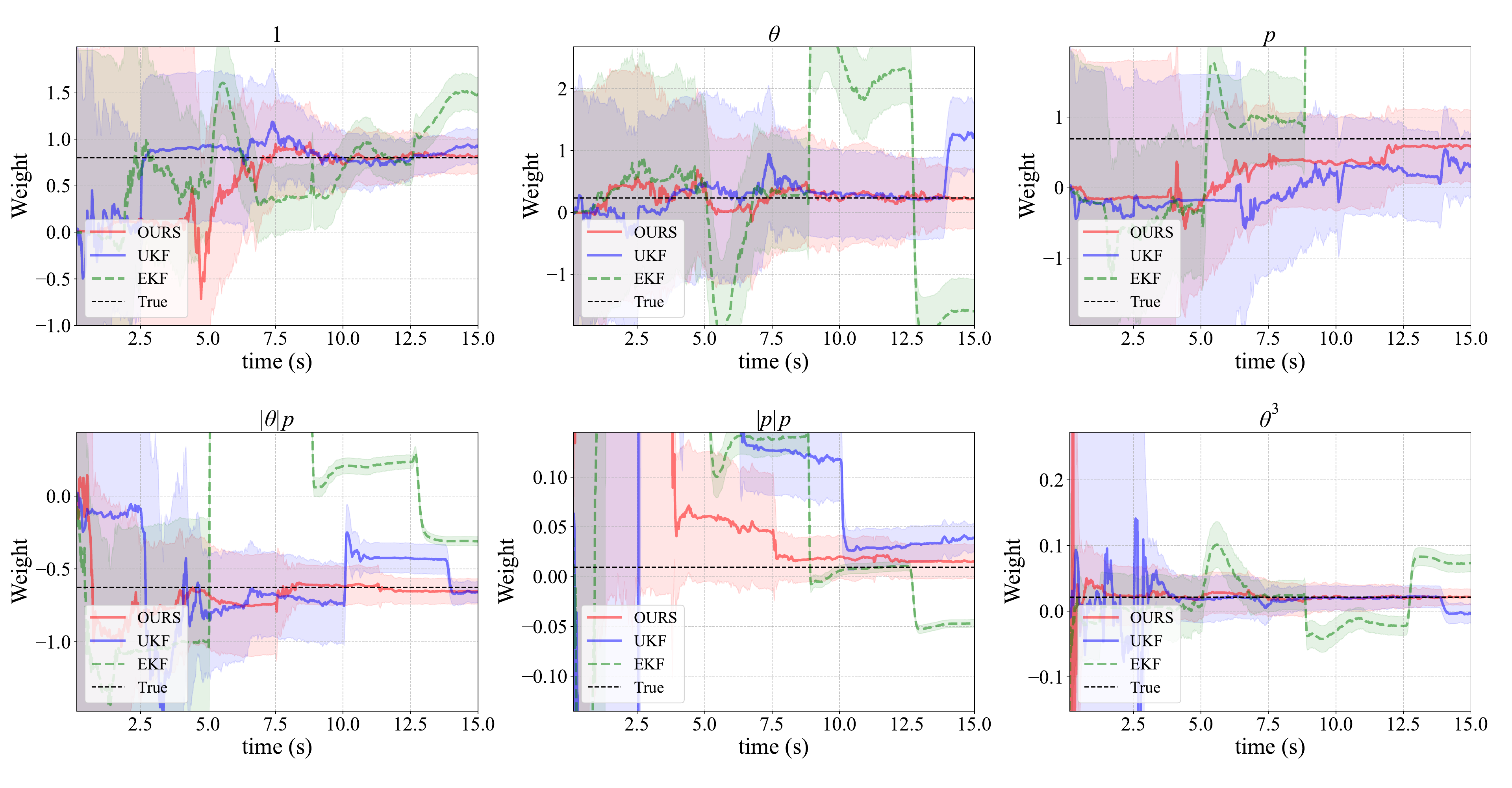}
  \caption{Estimation results for the coefficients (weight parameters) and their associated 1.96 standard deviation intervals. The red curve denotes the proposed SKI method, the blue curve represents the augmented UKF, and the green curve indicates the augmented EKF. Due to the large estimation error of the SINDy algorithm, its results are omitted for clarity.}
  \label{fig:exp1_1}
\end{figure*}


As illustrated in Fig.~\ref{fig:exp1_1}, the experimental results provide a clear comparison of the parameter estimation performance among the evaluated algorithms. It is evident that, compared to the EKF, the UKF demonstrates improved capability in accommodating the nonlinear characteristics of the system, yielding faster convergence, enhanced stability, and higher estimation accuracy. This observation substantiates the rationale for adopting the augmented UKF as the core identification framework in this study, particularly for nonlinear dynamical models. Furthermore, the proposed SKI algorithm, which incorporates online ARD-based sparsification mechanism, achieves even greater stability and accuracy than the augmented UKF, as reflected by its faster convergence and narrower confidence intervals. These results collectively validate the sparsity-promoting effectiveness of the SKI method.

\begin{table}[h]
  \centering
  \caption{Estimation Accuracy and Computational Efficiency of Different Algorithms.}
  \begin{tabular}{cccccc}
    \toprule
  \textbf{Method} & \textbf{SINDy} & \textbf{Augmented-EKF} & \textbf{Augmented-UKF} & \textbf{Our Method} \\
    \midrule
  \textbf{Mean $\ell_1$ Error} & 11.45 & 5.85 & 0.95 & \textbf{0.15} \\
  \textbf{Average Runtime (ms)} & - & 1.54 & \textbf{1.18} & 2.49 \\
    \bottomrule
  \end{tabular}
  \label{tab:1}
\end{table}

Given that the true values of the WingRock system parameters are known, the identification accuracy of each algorithm is quantitatively evaluated using the mean $\ell_1$-norm error between the estimated and true coefficients. Additionally, the average computational time per step for each algorithm is reported to assess efficiency. As shown in Table~\ref{tab:1}, our SKI method significantly outperforms the baseline methods in terms of identification accuracy, achieving an $84.21\%$ improvement over the Augmented-UKF, while maintaining comparable computational millisecond-level efficiency. It can also be observed that the UKF is faster than the EKF, primarily due to the use of a Cholesky-based implementation that enables efficient parallel batch computations, whereas the EKF does not benefit from such optimizations. Furthermore, the EKF relies on automatic differentiation to compute the state transition Jacobian matrix, which further increases its computational cost. As for the SINDy algorithm, since it is a non-recursive algorithm that requires all historical data for batch training, it is not meaningful to directly compare the single-step training time of SINDy with the algorithms under the AKF framework. Therefore, the per-step computation time of SINDy is not reported here.


\begin{figure*}[h]
  \centering
  \includegraphics[width=1.0\textwidth]{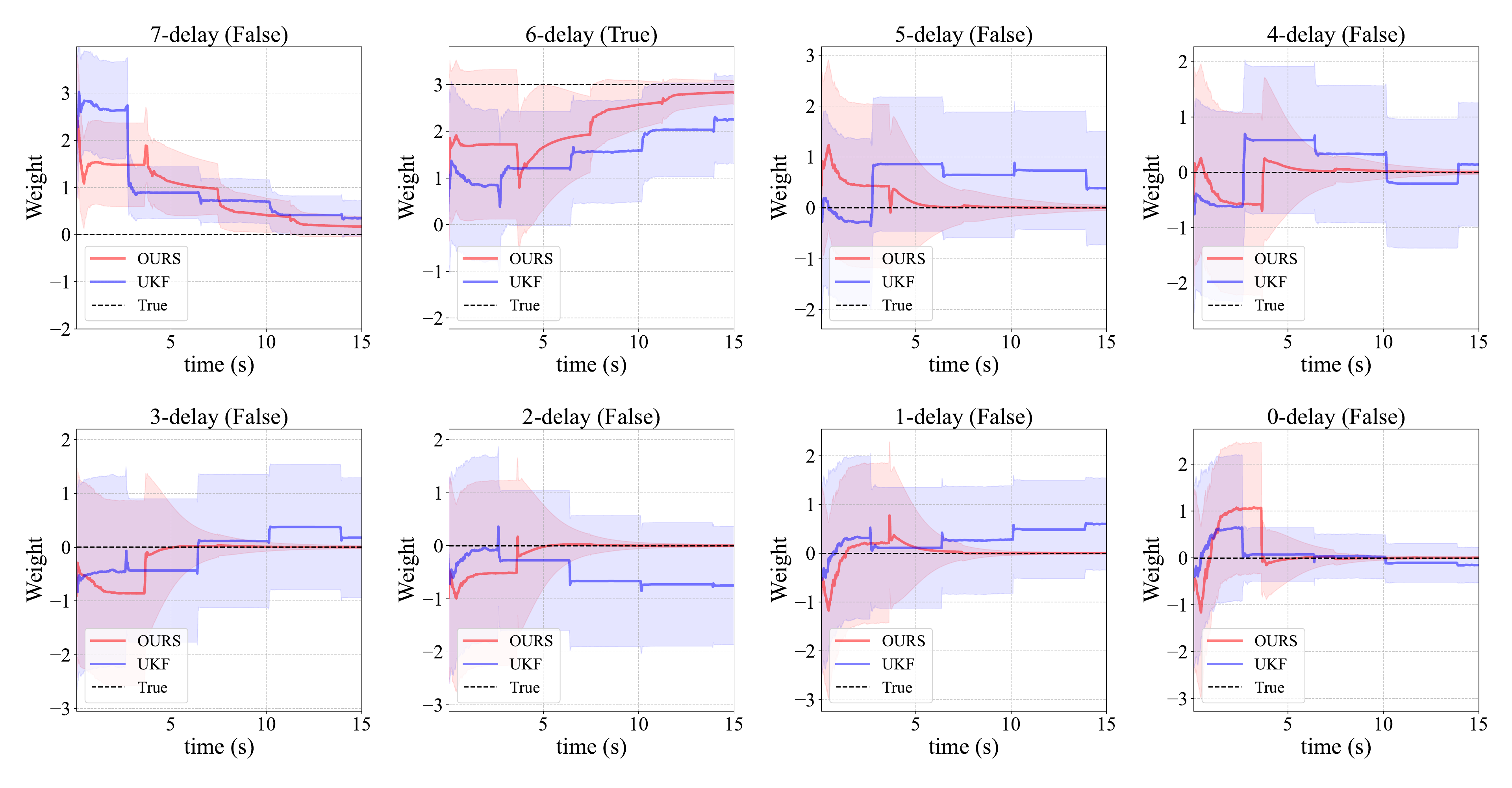}
  \caption{Estimated input gain trajectories for each candidate delay. Blue: Augmented UKF; Red: the proposed SKI method.}
  \label{fig:exp2_1}
\end{figure*}

\begin{figure}[h]
  \centering
  \includegraphics[width=0.84\textwidth]{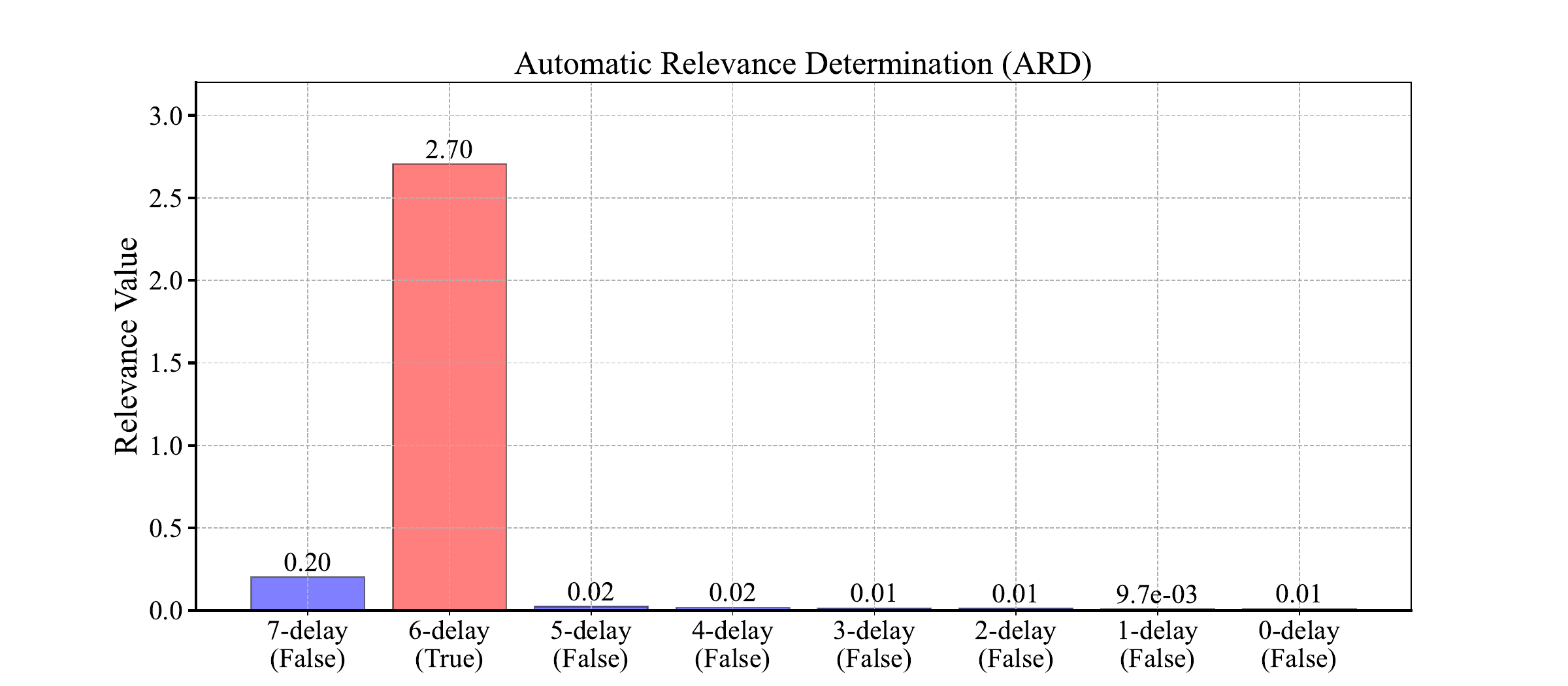}
  \caption{Identification results for different candidate time delays of the Online-ARD algorithm. Higher relevance indicates a more probable true delay.}
  \label{fig:exp2_2}
\end{figure}

To further illustrate the sparse identification capability of the SKI method, we design a dedicated time-delay identification experiment based on the WingRock system. In this task, the control input gain $L$ is considered an unknown parameter, and an unknown input delay (true value: 6 steps) is introduced. To enable simultaneous estimation of both the control effectiveness and the input delay, a sliding window of length $D=8$ is used to record the history of control inputs. The control effect at time $t$ is thus modeled as a weighted sum of the past $D$ inputs:
\begin{equation}
    \dot{p}_t = \sum_{j=1}^{D} L_j\,\Delta d_{t-j+1} + \Delta(\theta_t, p_t)
    \label{eq:delay}
\end{equation}
where $\bar{\bm{L}} = [L_1, L_2, \ldots, L_8]^{T}$ is the vector of candidate input gains, $\Delta d_{t-j+1}$ is the control input at time $t-j+1$, and $\Delta(\theta_t, p_t)$ is assumed known for this experiment. In this formulation, each delayed input acts as a candidate basis function, and the ARD-based algorithm adaptively determines the relevance of each delay by promoting sparsity in $\bar{\bm{L}}$. The most certain delay is identified as the one with the largest estimated gain.

For comparison, only the Augmented UKF is used as a baseline, since SINDy and Augmented EKF have been shown to perform poorly in this context (see Table~\ref{tab:1}). As shown in Fig.~\ref{fig:exp2_1}, the proposed SKI method rapidly and unambiguously identifies the true input delay (6 steps) and achieves more accurate estimation of the input gain compared to the Augmented UKF. Quantitatively, the proposed method yields an $\ell_1$ relative error for $L$ of \textbf{0.03}, which is lower than that of the Augmented UKF (\textbf{0.41}), demonstrating superior identification accuracy. Moreover, Fig.~\ref{fig:exp2_2} illustrates the final prior variances assigned to each candidate delay by the Online-ARD algorithm. The basis function (delayed inputs) with the largest prior variance (indicated by the red bar) corresponds to the most relevant feature, with the true 6-step delay being unambiguously identified by its pronounced variance. This result demonstrates the algorithm's strong capability for interpretable and selective sparse structure identification.

\subsection{Quadrotor UAV Simulation Experiment}
\label{subsec:exp2}

\begin{figure*}[h]
  \centering
  \begin{subfigure}[b]{0.36\textwidth}
    \centering
    \includegraphics[width=\textwidth]{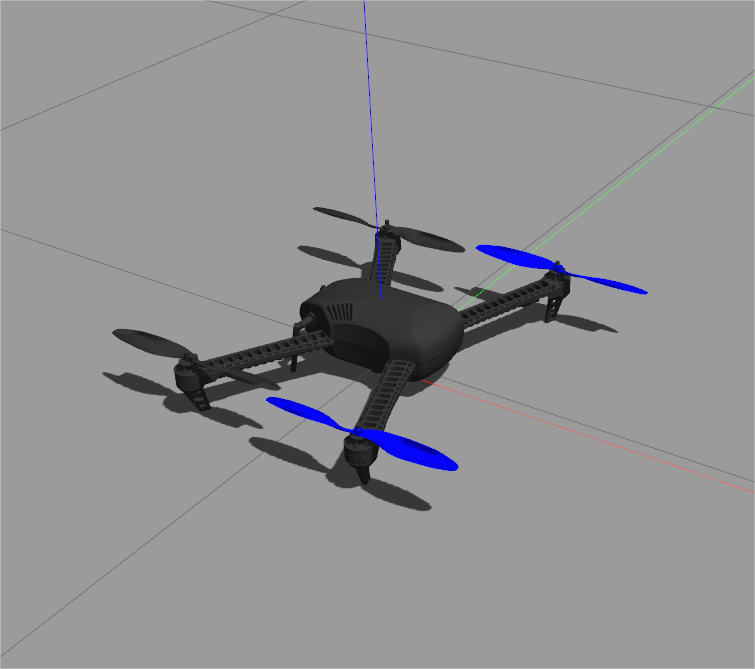}
    \caption{Simulated Quadrotor UAV}
    \label{subfig:sim_uav}
  \end{subfigure}
  \hspace{5mm} 
  \begin{subfigure}[b]{0.36\textwidth}
    \centering
    \includegraphics[width=\textwidth]{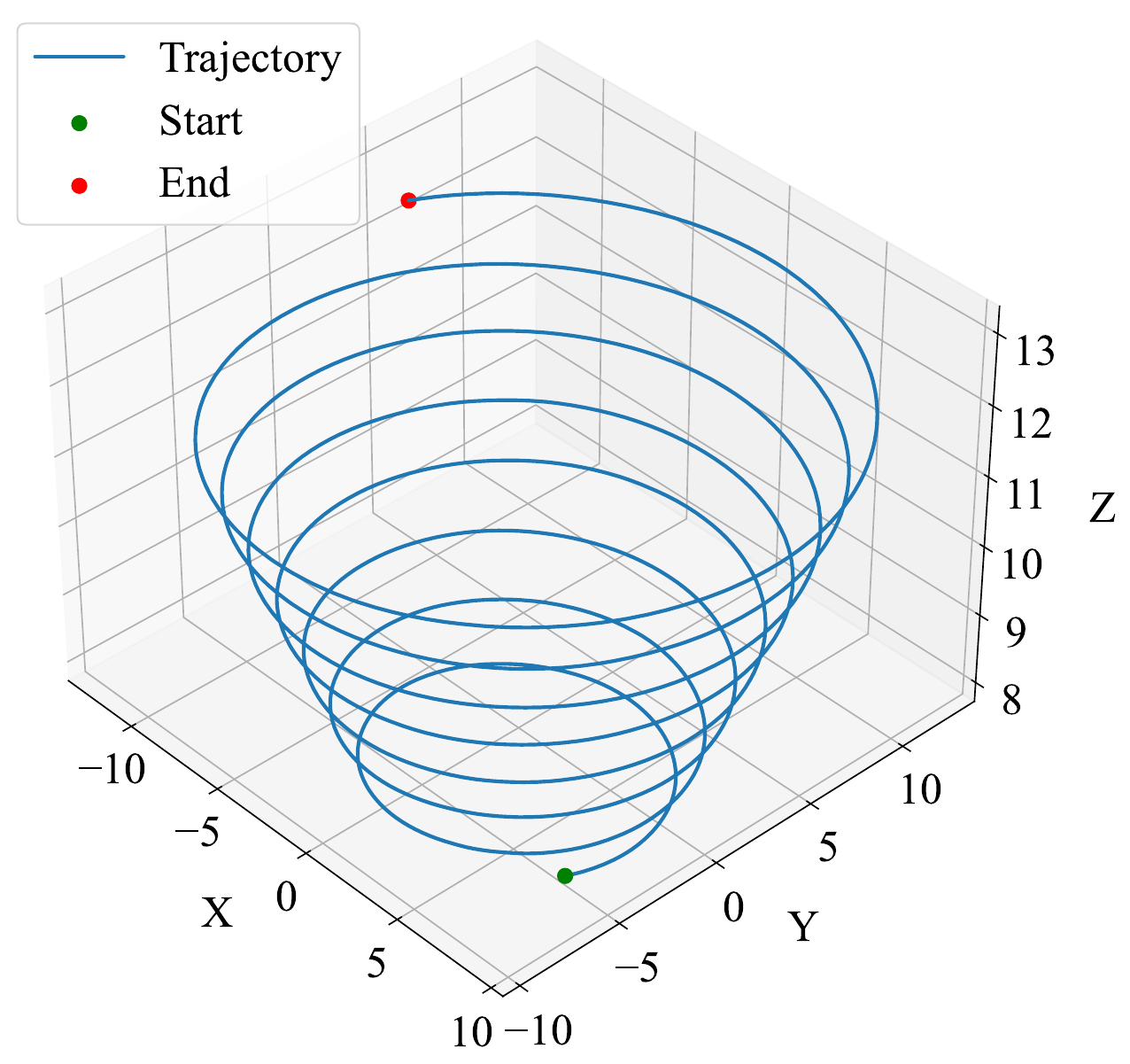}
    \caption{Flight Trajectory}
    \label{subfig:sim_traj}
  \end{subfigure}
  \caption{(a) The quadrotor UAV model in the Gazebo simulation environment; (b) the simulated spiral flight trajectory of the UAV.}
  \label{fig:sim_overview}
\end{figure*}

To rigorously evaluate the proposed SKI method under controlled and repeatable conditions, a high-fidelity simulation study was conducted using the Gazebo robotics simulation platform. The quadrotor UAV in this experiment is operated under the PX4 autopilot control algorithm, ensuring realistic and industry-standard flight dynamics and control behavior. During the simulation, the UAV is commanded to follow a three-dimensional spiral ascent trajectory, where the radius of the spiral increases linearly with time while the period of each revolution remains constant, as illustrated in Fig.~\ref{subfig:sim_traj}. This trajectory design ensures persistent excitation of the UAV's translational dynamics in all three spatial dimensions, thereby facilitating robust and comprehensive parameter identification.

In the simulated UAV dynamics, the thrust generated by the rotors is parameterized as a polynomial function of the average PWM input, with the true underlying relationship being approximately linear. The primary identification objective is to elucidate the functional mapping between the total thrust acceleration acting on the UAV and the average PWM input, thereby determining whether this relationship is best characterized by a linear, quadratic, or higher-order polynomial model. In addition to thrust modeling, both linear and quadratic aerodynamic drag effects are incorporated, with their respective weight parameters treated as unknown parameters to be identified. The magnitude of the drag force is proportional to the speed (i.e., the Euclidean norm of the velocity vector) and is always oriented opposite to the direction of motion. The state vector in the simulation is defined as $\bm{x} = [p^x, p^y, p^z, v^x, v^y, v^z]^{T}$, where $\bm{p} = [p^x, p^y, p^z]^{T}$ and $\bm{v} = [v^x, v^y, v^z]^{T}$ denote the UAV's position and velocity in the world coordinate frame, respectively. The control input comprises the PWM signals and the UAV's attitude (represented as a quaternion), while the model parameters to be identified include the thrust and drag coefficients. The continuous-time state transition model is formulated as follows:
\begin{equation}
    \begin{aligned}
        \dot{\bm{p}}_t &= \bm{v}_t \\
        \dot{\bm{v}}_t &= \bm{R}(\bm{q}_t) \bm{a}_t^{\text{body}} + \bm{a}_t^{\text{drag}}(\bm{v}_t) + \bm{g}
    \end{aligned}
    \label{eq:sim_3d_dynamics}
\end{equation}
where $\bm{p}_t = [p^x, p^y, p^z]^{T}$ is the position, $\bm{v}_t = [v^x, v^y, v^z]^{T}$ is the velocity, $\bm{q}_t$ is the attitude quaternion, $\bm{R}(\bm{q}_t)$ is the rotation matrix from body to world frame, $\bm{a}_t^{\text{body}} = [0, 0, a_t^{\text{thrust}}]^{T}$ is the thrust-induced acceleration in the body frame, $\bm{a}_{\text{drag}}(\bm{v}_t)$ is the aerodynamic drag acceleration, and $\bm{g} = [0, 0, -g]^{T}$ is the gravitational acceleration.

The thrust $a_t^{\text{thrust}}$ is parameterized as a polynomial function of the PWM input:
\begin{equation}
  \begin{aligned}
    a_t^{\text{thrust}} &= \sum_{i=0}^{5} w_i \left( \frac{1}{4} \sum_{j=1}^{4} (\mathrm{PWM}_t^j)^i \right)\\
    &= \sum_{i=0}^{5} w_i \mathrm{PWM}^i_{\mathrm{avg}}
    \label{eq:thrust_poly}
  \end{aligned}
\end{equation}
where $w_i$ are the weight parameters to be identified, and $\mathrm{PWM}_t^j$ denotes the PWM value of the $j$-th motor at time $t$. 

The aerodynamic drag acceleration is modeled as
\begin{equation}
    \bm{a}_{\text{drag}}(\bm{v}) = -d_1 \bm{v} - d_2 \|\bm{v}\|\bm{v}
    \label{eq:drag_poly}
\end{equation}
where $d_1$ and $d_2$ are the linear and quadratic drag coefficients, and $\|\bm{v}\|$ denotes the Euclidean norm of the velocity vector. Notably, the drag is assumed to be independent of the UAV's attitude and acts solely in the direction opposite to the velocity vector. To reduce the correlation among polynomial basis functions of different orders, the PWM input data were preprocessed by centering (i.e., removing the acceleration required for hovering) and standardization, which also increases the importance of the constant term in the basis expansion.

\begin{figure*}[h]
  \centering
  \includegraphics[width=1.0\textwidth]{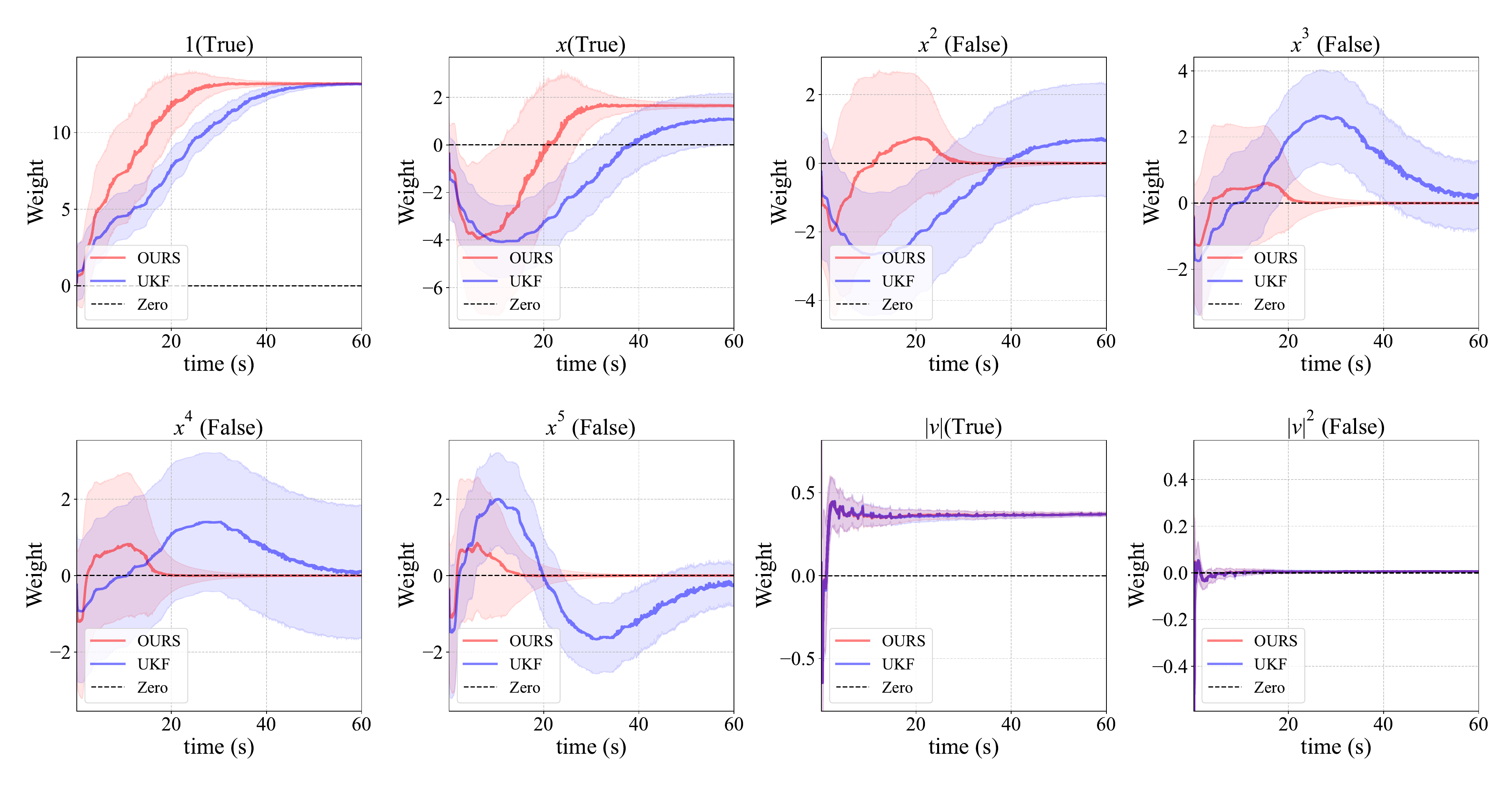}
  \caption{Estimated trajectories of the coefficients (weight parameters) for each candidate PWM basis function and the linear/quadratic drag terms. Blue: Augmented UKF; Red: the proposed SKI method.}
  \label{fig:exp3_1}
\end{figure*}
\begin{figure}[h]
  \centering
  \includegraphics[width=0.84\textwidth]{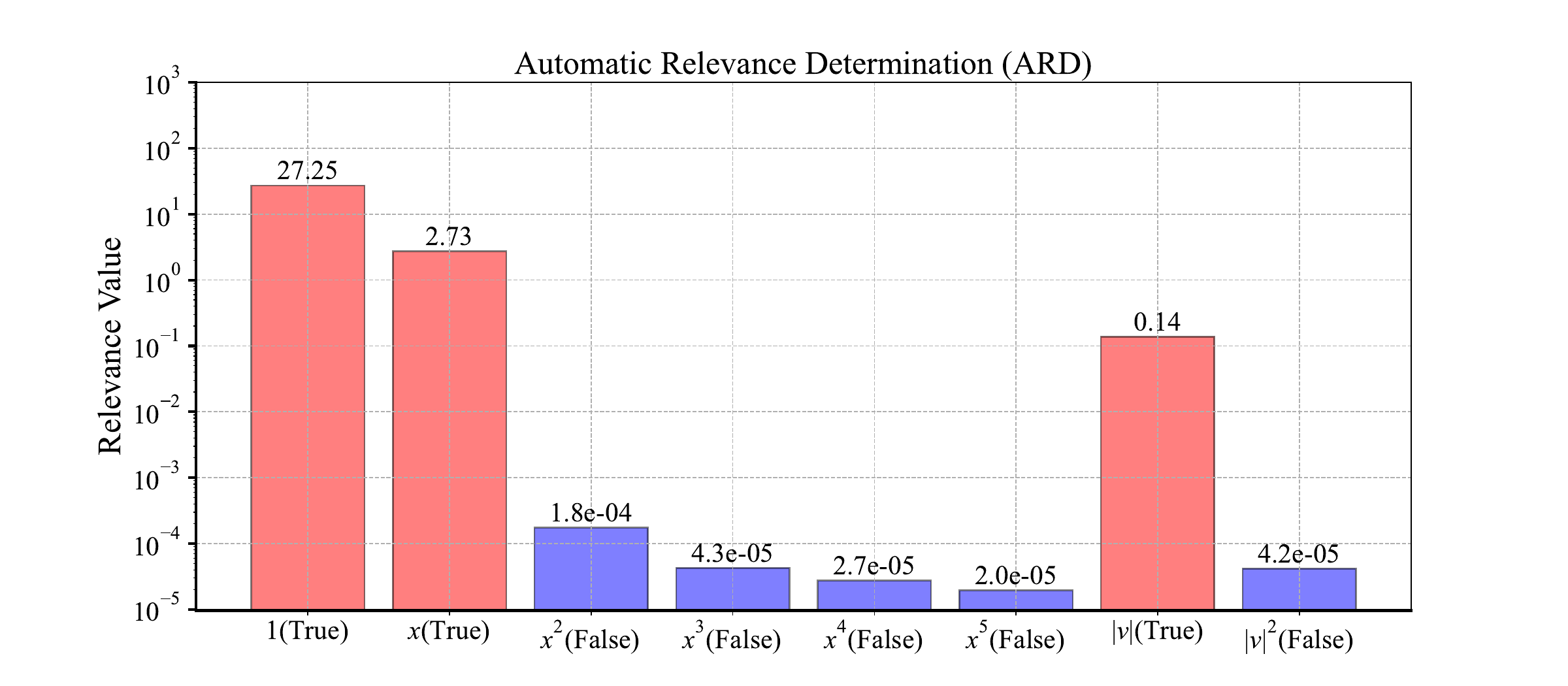}
  \caption{Identification results for the PWM basis functions and the linear/quadratic drag terms using the Online-ARD algorithm. The effectively selected terms are highlighted with red bars. Note: the y-axis is in logarithmic scale, as the prior variances of the selected terms differ by several orders of magnitude; the log scale facilitates clear comparison.}
  \label{fig:exp3_2}
\end{figure}

The complete set of candidate basis functions for the identification task is thus defined as:
\begin{equation}
\begin{aligned}
\Phi(\mathrm{PWM}_t^{\mathrm{avg}}, \|\bm{v}_t\|) 
&= \big[\, 1,\; \mathrm{PWM}_t^{\mathrm{avg}},\; (\mathrm{PWM}_t^{\mathrm{avg}})^2,\; (\mathrm{PWM}_t^{\mathrm{avg}})^3, \\
&\quad\;\; (\mathrm{PWM}_t^{\mathrm{avg}})^4,\; (\mathrm{PWM}_t^{\mathrm{avg}})^5,\; \|\bm{v}_t\|,\; \|\bm{v}_t\|^2 \,\big]
\end{aligned}
\label{eq:sim_basis_functions_3d}
\end{equation}
where the first six terms correspond to the constant and polynomial terms of the (normalized) PWM input (thrust basis), and the last two terms represent the linear and quadratic speed-dependent drag components. The identification process involves estimating the weight vector $\bm{W} = [w_0, w_1, w_2, w_3, w_4, w_5, d_1, d_2]^{T}$ associated with the above basis functions. Ideally, the algorithm should primarily select the constant term ($w_0$), the linear PWM term ($w_1$), and the linear speed drag term ($d_1$), while suppressing the remaining coefficients, thus reflecting the true underlying physical relationships. The observation model is defined as a partial observation of the system state, where only the UAV's three-dimensional position in the world coordinate frame is available for measurement:
\begin{equation}
    \bm{y}_t = \bm{C} \bm{x}_t
    \label{eq:sim_observation}
\end{equation}
where $\bm{H} \in \mathbb{R}^{3 \times 6}$ is the observation matrix selecting the position components, i.e., $\bm{H} = [\bm{I}_3,\, \bm{0}_{3 \times 3}]$. It is further assumed that the drag experienced by the UAV is independent of its attitude.

In the simulation experiment, the observation data were collected at a frequency of 50~Hz. The total duration of the experiment was 60~s. The baseline algorithm for comparison remained the Augmented UKF. The experimental results are presented in Fig.~\ref{fig:exp3_1} and Fig.~\ref{fig:exp3_2}. It is evident that only the constant term, the linear PWM term, and the linear drag term were distinctly identified by the proposed SKI method, which aligns with the underlying UAV dynamics model established in Gazebo prior to the simulation experiments. Furthermore, a comparative analysis reveals that, relative to the Augmented UKF, our algorithm demonstrates advantages in terms of stability and sparsity. These results substantiate the feasibility and enhanced effectiveness of the proposed method for UAV dynamics identification tasks. 

\subsection{Physical UAV Flight Experiment}
\label{subsec:exp3}

\begin{figure*}[h]
  \centering
  \begin{subfigure}[b]{0.31\textwidth}
    \centering
    \includegraphics[width=\textwidth]{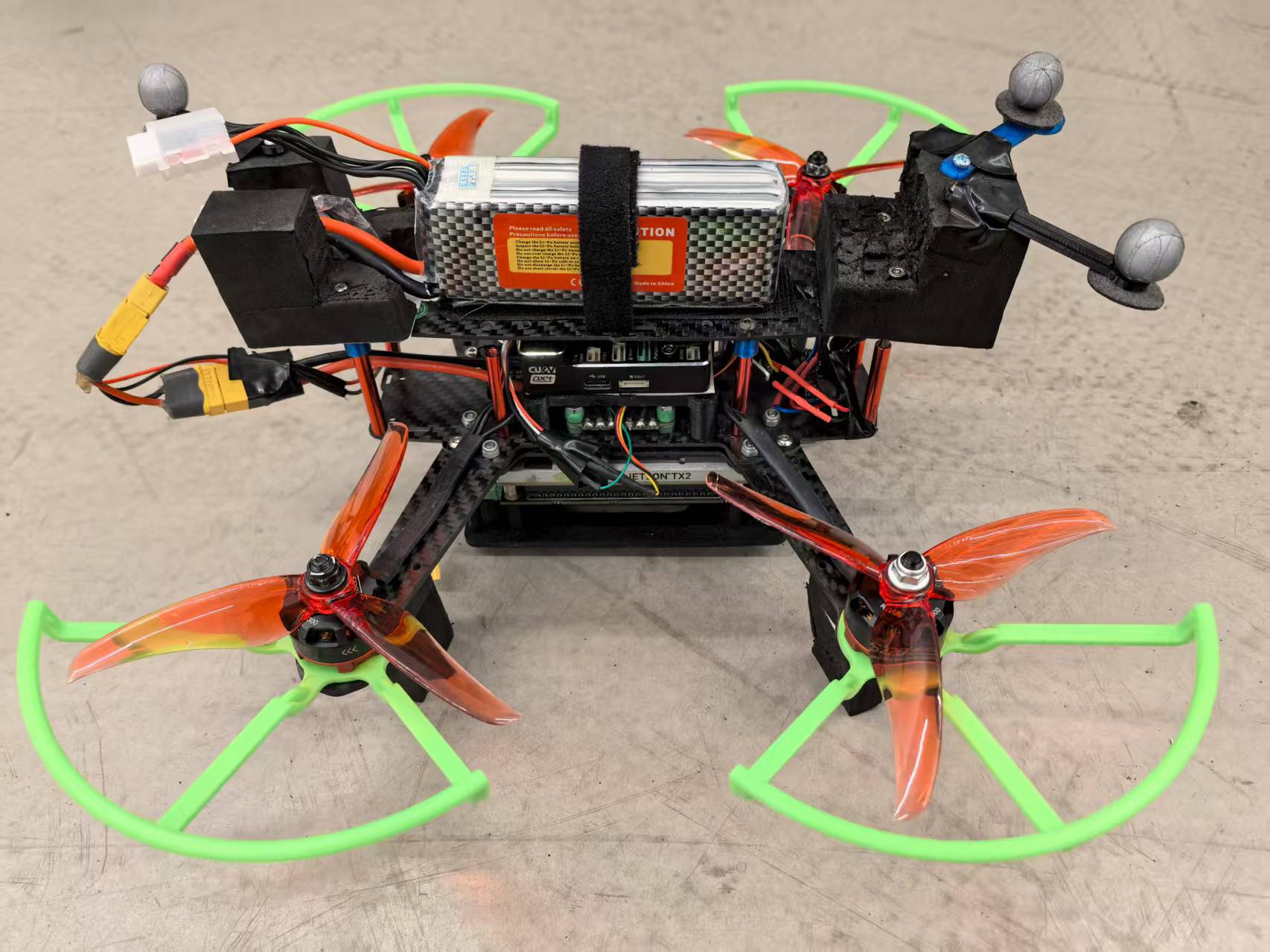}
    \caption{Quadrotor UAV}
    \label{subfig:phys_uav}
  \end{subfigure}
  \hfill
  \begin{subfigure}[b]{0.31\textwidth}
    \centering
    \includegraphics[width=\textwidth]{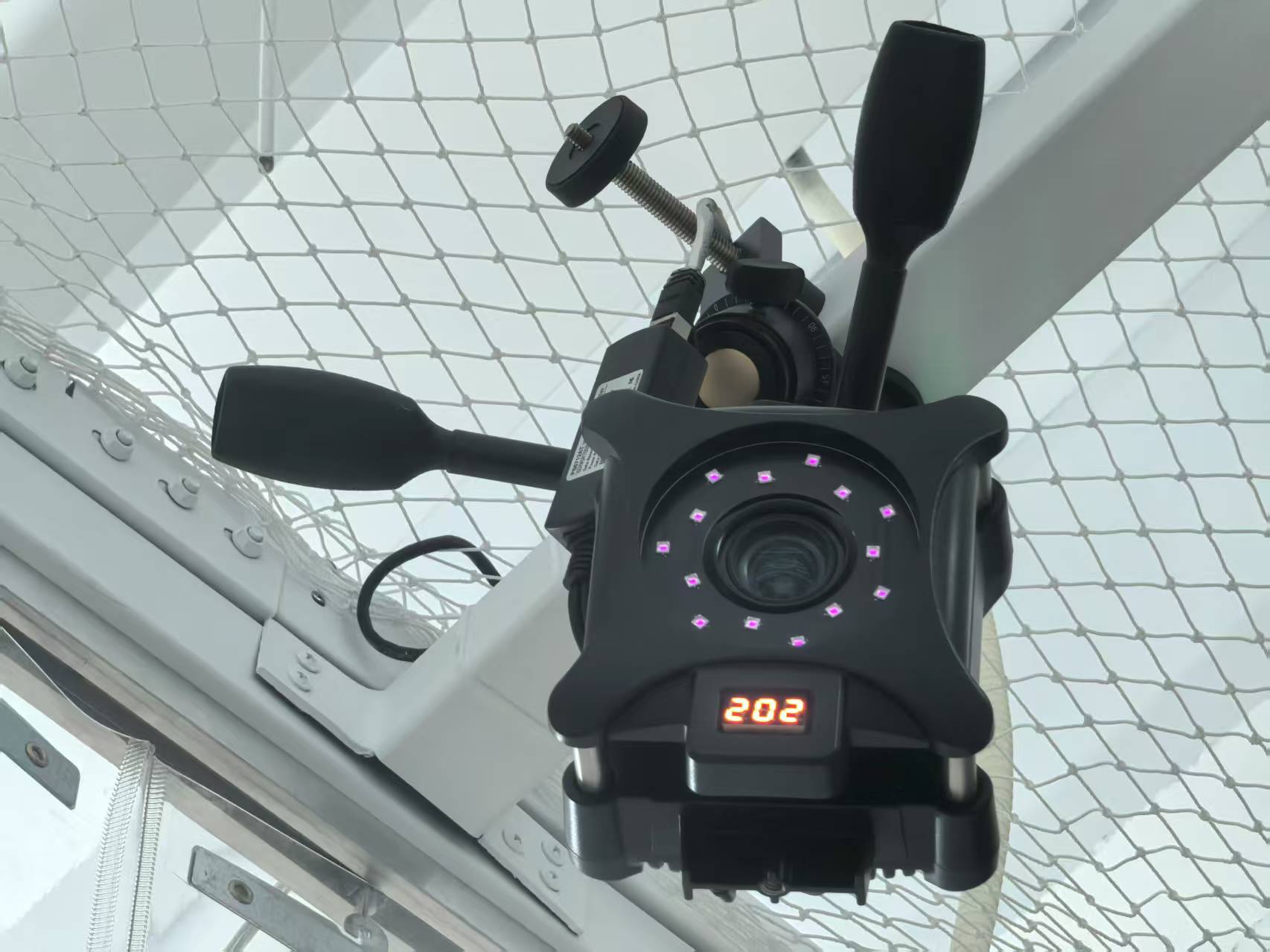}
    \caption{Motion Capture System}
    \label{subfig:phys_mocap}
  \end{subfigure}
  \hfill
  \begin{subfigure}[b]{0.347\textwidth}
    \centering
    \includegraphics[width=\textwidth]{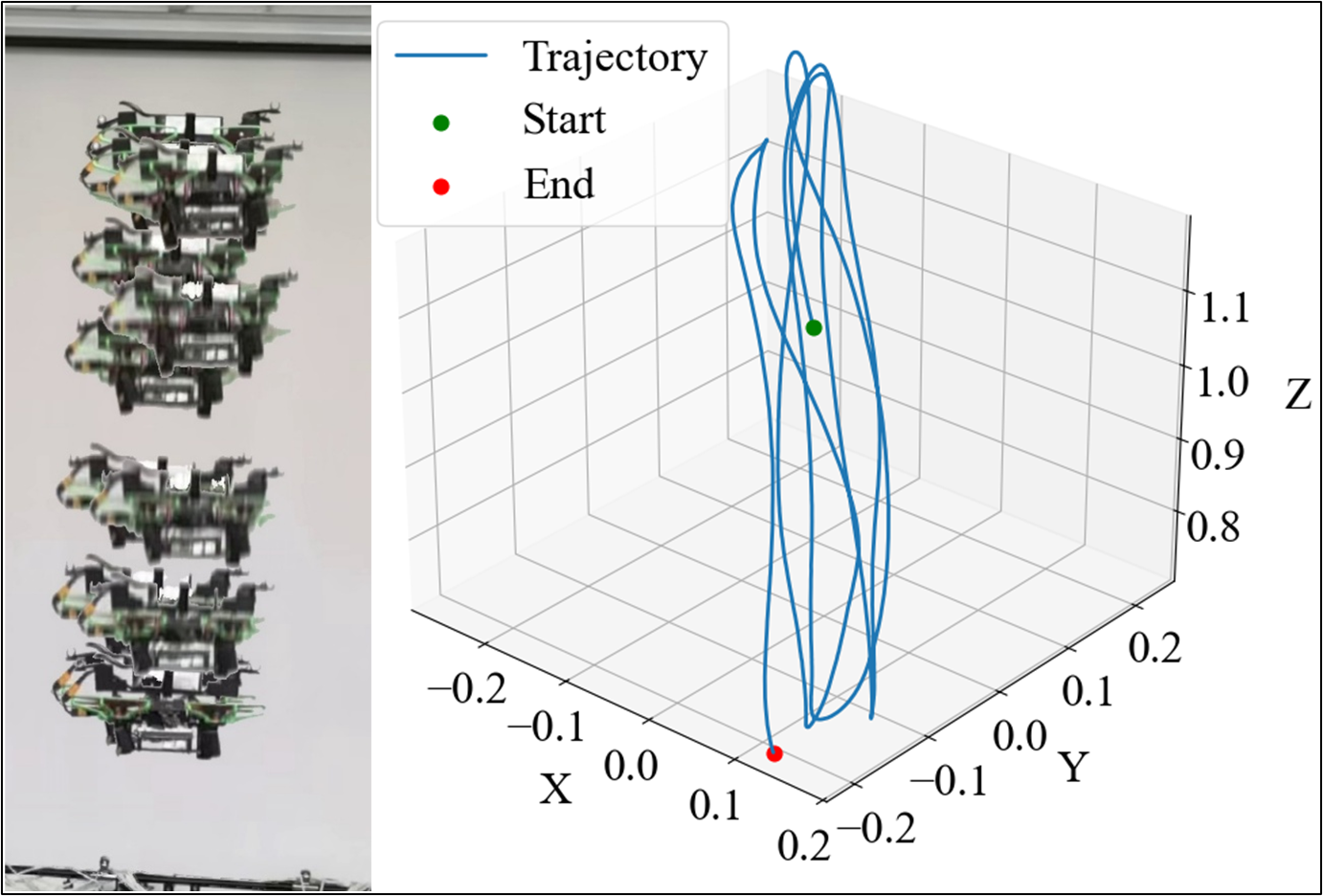}
    \caption{Flight Trajectory}
    \label{subfig:phys_traj}
  \end{subfigure}
  \caption{(a) The quadrotor UAV used in the physical experiment; (b) motion capture system for measuring UAV position in the world frame; (c) schematic of the UAV body frame and the world frame defined by the motion capture system.}
  \label{fig:phys_overview}
\end{figure*}

To further assess the efficacy of the proposed SKI method in real-world scenarios, a physical flight experiment was conducted using a quadrotor UAV platform, as shown in Fig.~\ref{subfig:phys_uav}. The UAV is equipped with a PX4 flight controller, electronic speed controllers (ESCs), and a Jetson TX2 onboard computer. State measurements were obtained via a NOKOV-MARS motion capture system (see Fig.~\ref{subfig:phys_mocap}), which provides three-dimensional position data in the world coordinate frame at a frequency of 100~Hz.

In this experiment, the UAV was commanded to follow a vertical oscillatory trajectory along the $z$-axis, characterized by a sinusoidal pattern, as depicted in Fig.~\ref{subfig:phys_traj}. This trajectory design ensures persistent excitation of the vertical dynamics, thereby enabling accurate identification of the mapping between the average PWM input and the vertical acceleration.

It is important to note that, due to the limited size of the experimental site, the UAV's flight speed remained relatively low throughout the experiment. Consequently, aerodynamic drag effects were negligible and thus not considered in the identification process. The identification task in the physical experiment was therefore restricted to the basis functions associated with the PWM input, without the inclusion of drag-related terms. This stands in contrast to the simulation study.

The system dynamics were modeled analogously to the simulation case, but restricted to the vertical direction. The state vector is defined as $\bm{x} = [z, v^z]^{T}$, where $z$ denotes the vertical position and $v^z$ the vertical velocity in the world frame. The control input is the average PWM signal, $\mathrm{PWM}_t^{\mathrm{avg}}$, across all four motors. The continuous-time state-space model is given by:
\begin{equation}
    \begin{aligned}
        \dot{z}_t &= v_t^z \\
        \dot{v}_t^z &= a_t^{\text{thrust}} - g
    \end{aligned}
    \label{eq:uav_dynamics}
\end{equation}
where $a_t^{\text{thrust}}$ denotes the thrust-induced acceleration along the $z$-axis, and $g$ is the gravitational acceleration. The observation model is a partial observation of the state, with only the vertical position $z_t$ being measured.

The candidate basis functions for the identification task in the physical experiment are thus defined as:
\begin{equation}
    \bm{\Phi}(\mathrm{PWM}_t^{\mathrm{avg}}) = \left[\, 1,\; \mathrm{PWM}_t^{\mathrm{avg}},\; (\mathrm{PWM}_t^{\mathrm{avg}})^2,\; (\mathrm{PWM}_t^{\mathrm{avg}})^3,\; (\mathrm{PWM}_t^{\mathrm{avg}})^4,\; (\mathrm{PWM}_t^{\mathrm{avg}})^5 \,\right]
    \label{eq:real_basis_functions}
\end{equation}
where the terms correspond to the constant and polynomial terms of the (normalized) PWM input up to the fifth order. The identification process involves estimating the weight vector $\bm{w} = [w_0, w_1, w_2, w_3, w_4, w_5]^{T}$ associated with these basis functions.

The mapping from the average PWM input to the vertical acceleration is thus parameterized as:
\begin{equation}
    a^{\text{thrust}}_t = \sum_{i=0}^{5} w_i \, (\mathrm{PWM}^{\mathrm{avg}}_t)^i
    \label{eq:thrust_poly_real}
\end{equation}
where $w_i$ are the coefficients (weight parameters) to be identified. 

\begin{figure*}[htbp]
  \centering
  \includegraphics[width=1.0\textwidth]{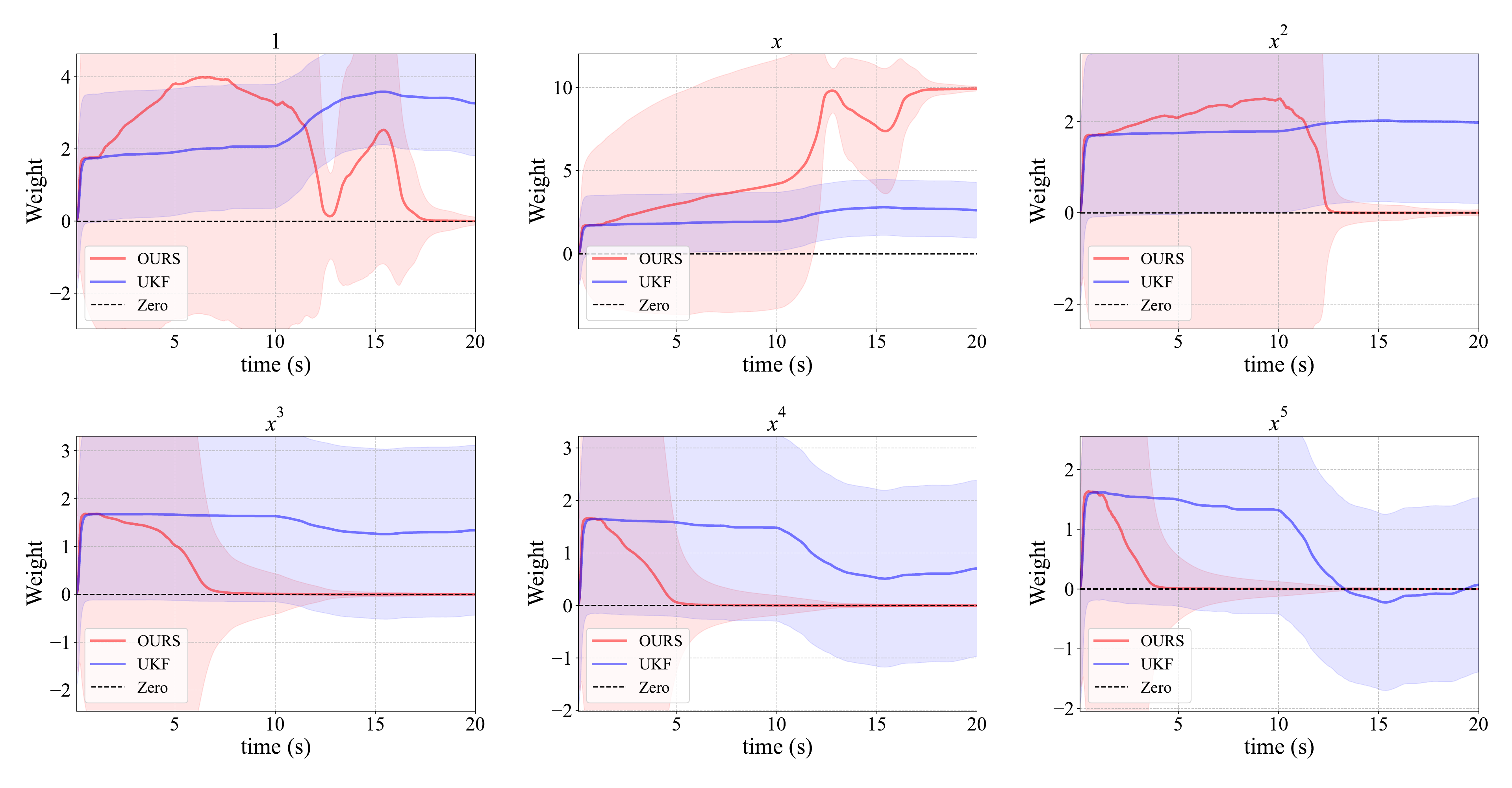}
  \caption{Comparison of the identification and selection of PWM basis functions by the Augmented-UKF (blue) and the proposed SKI method (red) in the physical quadrotor experiment.}
  \label{fig:exp4_1}
\end{figure*}

\begin{figure}[H]
  \centering
  \includegraphics[width=0.84\textwidth]{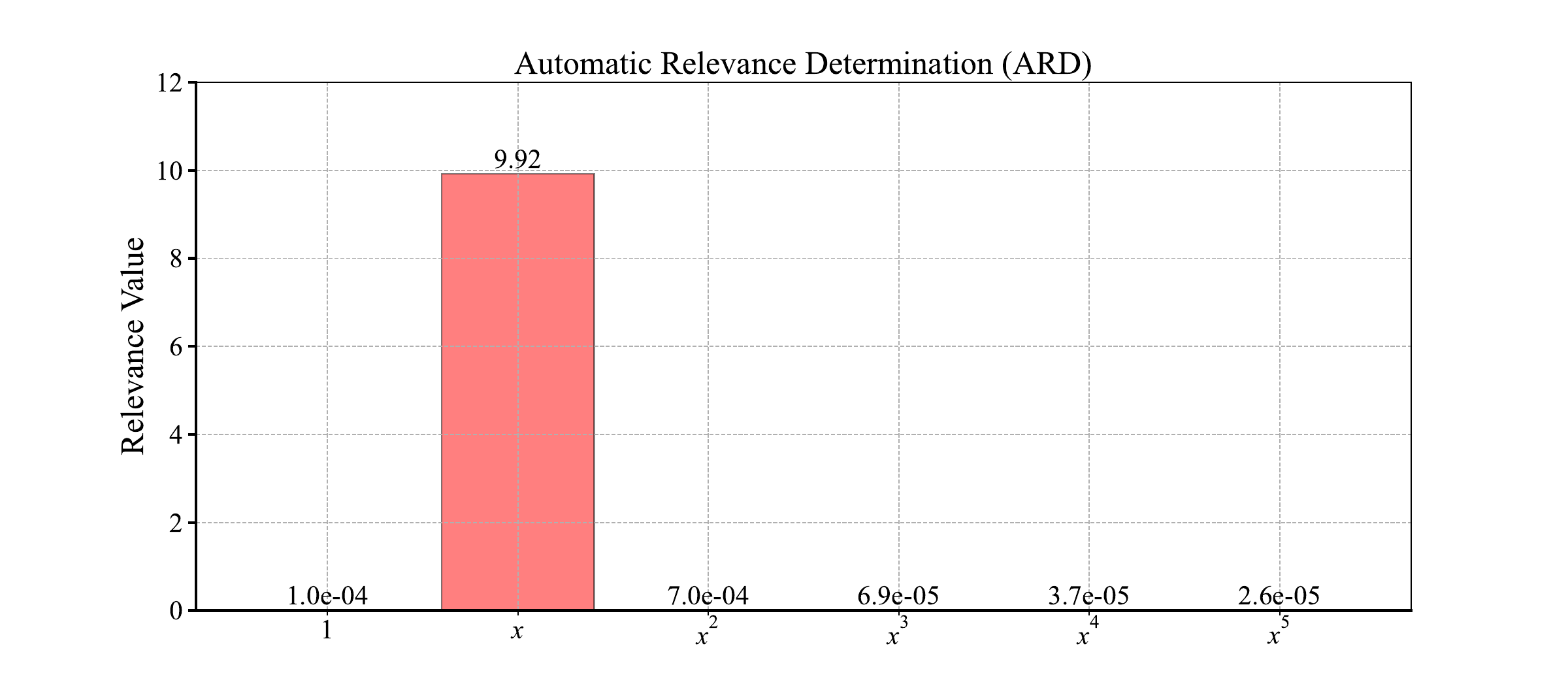}
  \caption{Identification results for the PWM basis functions using the Online-ARD algorithm in the physical flight experiment. The effectively selected terms are highlighted with red bars.}
  \label{fig:exp4_2}
\end{figure}

As shown in Fig.~\ref{fig:exp4_1} and Fig.~\ref{fig:exp4_2}, the identification results from the physical experiment are consistent with those from the simulation, with the linear term again identified as dominant. Unlike the simulation, the real-world data were collected online without centering or preprocessing, so the constant term (related to hovering acceleration) is absorbed into the linear term and not selected separately. The proposed Sparse Kalman Identification (SKI) method maintains better stability and sparsity than the Augmented UKF, confirming its practical effectiveness.

\section{Conclusion}
\label{sec:conclusion}
In summary, this work proposes a Sparse Kalman Identification (SKI) method for real-time sparse identification of nonlinear dynamics, which can achieve accurate, sparse, and interpretable parameter estimation from noisy and partial observations. The key innovation is the incorporation of the ARD algorithm, which enables online estimation of the prior variances associated with each candidate basis function weight. This adaptive mechanism allows the method to dynamically select relevant features and suppress irrelevant ones throughout the sparse identification process. As a result, the SKI method enhances model interpretability and improves robustness against overfitting and redundancy in the candidate basis functions. Comprehensive simulations and real-world flight experiments were conducted to evaluate the proposed SKI method. The results show that SKI consistently outperforms conventional methods in sparsity and accuracy, while maintaining competitive computational efficiency. These findings demonstrate the practical utility and robustness of SKI for engineering applications.

\section{Acknowledgement}
\label{sec:acknowledgement}

This work was supported by the National Natural Science Foundation of China (Grant No. 12572052) and the Beijing Natural Science Foundation (Grant No. L251013). The authors gratefully acknowledge the financial support provided by these funding agencies, which made this research possible.

\section*{Appendix}

\subsection{Cholesky Decomposition in the UKF Sigma Point Generation}
\label{appendix:ukf_cholesky}

The standard UKF propagates and updates the full state covariance matrix, which can suffer from numerical instability. To improve computational efficiency and robustness, the square-root formulation of the UKF (SR-UKF) maintains and propagates the Cholesky factor of the covariance matrix directly, employing numerically stable linear algebraic operations such as Cholesky and QR decomposition. 

\paragraph{Initialization.} 
The algorithm is initialized by computing the Cholesky factor of the initial augmented covariance matrix:
\begin{equation}
\begin{aligned}
\bm{\Sigma}_0 &= 
\begin{bmatrix}
\bm{P}_0 & \bm{0} \\
\bm{0} & \bm{S}_0
\end{bmatrix} \\
\bm{\Sigma}_0 &= \bm{U}_0 \bm{U}_0^{T} \\
\bm{U}_0 &= \mathrm{chol}(\bm{\Sigma}_0)
\end{aligned}
\end{equation}
where $\bm{U}_0$ is the \textbf{lower-triangular} Cholesky factor of $\bm{\Sigma}_0$.

\paragraph{Sigma-point generation.} 
At time step $t-1$, given the posterior mean $\bm{\xi}_{t-1}$ and Cholesky factor $\bm{U}_{t-1}$, the sigma points are generated as:
\begin{equation}
\begin{aligned}
\bm{\chi}_{0,t-1} &= \bm{\xi}_{t-1}, \\
\bm{\chi}_{i,t-1} &= \bm{\xi}_{t-1} + \sqrt{L+\lambda}\,[\bm{U}_{t-1}]_{:,i}, \quad i=1,\ldots,L\\
\bm{\chi}_{i+L,t-1} &= \bm{\xi}_{t-1} - \sqrt{L+\lambda}\,[\bm{U}_{t-1}]_{:,i}, \quad i=1,\ldots,L
\end{aligned}
\end{equation}
where $L$ is the dimension of the augmented state and $\lambda=L(\alpha^2-1)$ is the UKF scaling parameter.

\paragraph{Weights.}
The corresponding mean and covariance weights are defined as:
\begin{equation}
\begin{aligned}
W_0^{(m)} &= \frac{\lambda}{L+\lambda}, \\
W_0^{(c)} &= \frac{\lambda}{L+\lambda} + (1-\alpha^2+\beta),\\
W_i^{(m)} &= W_i^{(c)} = \frac{1}{2(L+\lambda)}, \quad i=1,\ldots,2L
\end{aligned}
\end{equation}
where $\alpha$ determines the spread of the sigma points and $\beta$ incorporates prior distribution knowledge (for Gaussian distributions, $\beta=2$).

\paragraph{Time update.}
Each sigma point is propagated through the nonlinear state transition function:
\begin{equation}
\bm{\chi}_{i,t|t-1} = \bm{\bar{F}}(\bm{\chi}_{i,t-1}, \bm{u}_{t-1}), \quad i=0,\ldots,2L
\end{equation}
The predicted mean is obtained as a weighted sum:
\begin{equation}
\bm{\xi}_t^- = \sum_{i=0}^{2L} W_i^{(m)} \bm{\chi}_{i,t|t-1}
\end{equation}
The predicted Cholesky factor of the covariance is computed via QR decomposition:
\begin{equation}
\begin{aligned}
\bm{A}_x &= \left[ \sqrt{W_1^{(c)}}(\bm{\chi}_{1:2L,t|t-1}-\bm{\xi}_t^-),\; \bm{Q}^{1/2} \right],\\
\bm{A}_x &= \mathcal{Q}_x \mathcal{R}_x, \qquad \bm{U}_t^- = \mathcal{R}_x^{T}
\end{aligned}
\end{equation}
where $\bm{Q}^{1/2}$ is the lower-triangular Cholesky factor of the process noise covariance $\bm{Q}$.
If $W_0^{(c)} \neq 0$, a subsequent Cholesky rank-one update or downdate is applied:
\begin{equation}
\bm{U}_t^- = 
\mathrm{cholupdate}\!\left( \bm{U}_t^-,\, \bm{\chi}_{0,t|t-1}-\bm{\xi}_t^-,\, W_0^{(c)} \right)
\end{equation}

\paragraph{Measurement update.}
Each predicted sigma point is mapped through the measurement function:
\begin{equation}
\bm{\gamma}_{i,t} = \bm{h}(\bm{\chi}_{i,t|t-1}), \quad i=0,\ldots,2L
\end{equation}
The predicted measurement mean and covariance factor are computed as:
\begin{align}
\bm{y}_t^- &= \sum_{i=0}^{2L} W_i^{(m)} \bm{\gamma}_{i,t},\\
\bm{A}_y &= \left[ \sqrt{W_1^{(c)}}(\bm{\gamma}_{1:2L,t}-\bm{y}_t^-),\; \bm{R}^{1/2} \right],\\
\bm{A}_y &= \mathcal{Q}_y \mathcal{R}_y, \qquad \bm{U}_{y_t} = \mathcal{R}_y^{T}
\end{align}
where $\bm{R}^{1/2}$ is the lower-triangular Cholesky factor of the measurement noise covariance $\bm{R}$.
If $W_0^{(c)} \neq 0$, perform a Cholesky update/downdate:
\begin{equation}
\bm{U}_{y_t} = 
\mathrm{cholupdate}\!\left( \bm{U}_{y_t},\, \bm{\gamma}_{0,t}-\bm{y}_t^-,\, W_0^{(c)} \right)
\end{equation}

The cross-covariance between the state and measurement is given by:
\begin{equation}
\bm{C}_t = \sum_{i=0}^{2L} W_i^{(c)} (\bm{\chi}_{i,t|t-1}-\bm{\xi}_t^-) (\bm{\gamma}_{i,t}-\bm{y}_t^-)^{T}
\end{equation}
The Kalman gain is obtained by solving two triangular systems:
\begin{equation}
  \bm{K}_t = \bm{C}_t (\bm{U}_{y_t} \bm{U}_{y_t}^{T})^{-1}
\end{equation}
which avoids explicit matrix inversion.

\paragraph{Posterior update.}
The augmented state mean and Cholesky factor are updated as:
\begin{align}
\bm{\xi}_t^+ &= \bm{\xi}_t^- + \bm{K}_t(\bm{y}_t - \bm{y}_t^-),\\
\bm{\Gamma}_t &= \bm{K}_t \bm{U}_{y_t},\\
\bm{U}_t^+ &= \mathrm{cholupdate}\!\left(\bm{U}_t^-, \bm{\Gamma}_t, -1\right)
\end{align}

In summary, the square-root UKF maintains all covariance matrices in their Cholesky factorized form throughout the prediction and update steps by employing QR decomposition and Cholesky updating. This approach ensures numerical stability and computational efficiency, particularly in real-time and high-dimensional filtering applications, and is fully consistent with the notation and algorithmic structure presented in the main text.

\subsection{Detailed Expansion and Derivation of the ARD Loss Function}
\label{appendix:ard_loss}

In this section, we provide a detailed derivation of the loss function $\mathcal{L}$ used for prior variance update, with particular emphasis on the explicit expansion of the Gaussian likelihood term and the step-by-step derivation of the components $\mathcal{L}_1$ and $\mathcal{L}_2$. We also discuss how the formulation avoids the pathological case where $\bm{S}_0^{\mathrm{new}} = \bm{S}_0^{\mathrm{old}}$, which would otherwise result in vanishing gradients.

The loss function is derived from the Eq.\eqref{eq.ard_loss_pri2}, where the prior covariance matrix is updated from $\bm{S}_0^{\mathrm{old}}$ to $\bm{S}_0^{\mathrm{new}}$. The relevant term can be written as:
\begin{equation}
  \begin{aligned}
      \mathcal{L} &= -\ell(\bm{S}_0^\mathrm{new}) \\
      &= -\log C -\log \mathcal{N}(\bm{\theta} \mid \bm{m}_t^{\mathrm{old}}, \bm{S}_t^{\mathrm{old}} + \Delta\bm{S}_0) 
  \end{aligned}
\end{equation}

Expanding the log-likelihood of the multivariate Gaussian and expand $\log C$ according to Eq.\eqref{eq.gaussian_composition}, we obtain:
\begin{equation}
    \begin{aligned}
        \mathcal{L} &= -\log((2\pi)^{n_\theta/2} \frac{|\bm{S}_0^\mathrm{old}|^{1/2} |\Delta \bm{S}_0|^{1/2}}{|\bm{S}_0^\mathrm{new}|^{1/2}})
        +\frac{1}{2} \log \left| (2\pi)^{n_\theta/2} (\bm{S}_t^{\mathrm{old}} + \Delta\bm{S}_0) \right| + \frac{1}{2} (\bm{m}_t^{\mathrm{old}})^{T} (\bm{S}_t^{\mathrm{old}} + \Delta\bm{S}_0)^{-1} \bm{m}_t^{\mathrm{old}} \\
        &= \frac{1}{2}\log|\bm{S}_0^\mathrm{new}| - \frac{1}{2} \log|\bm{S}_0^\mathrm{old}| - \frac{1}{2} \log|\Delta\bm{S}_0| + \frac{1}{2} \log \left| \bm{S}_t^{\mathrm{old}} + \Delta\bm{S}_0 \right| + \frac{1}{2} (\bm{m}_t^{\mathrm{old}})^{T} (\bm{S}_t^{\mathrm{old}} + \Delta\bm{S}_0)^{-1} \bm{m}_t^{\mathrm{old}}
    \end{aligned}
\end{equation}
where $n_\theta$ is the dimension of $\bm{\theta}$. In the context of optimization, the term $\frac{1}{2} \log|\bm{S}_0^\mathrm{old}|$ can be omitted, as it does not affect the gradient with respect to $\bm{S}_0^{\mathrm{new}}$, the loss can be decomposed as follows:
\begin{equation}
  \begin{aligned}
    \mathcal{L} &= \mathcal{L}_1 + \mathcal{L}_2,\\
        \mathcal{L}_1 &= \frac{1}{2} (\bm{m}_t^{\mathrm{old}})^{T} (\bm{S}_t^{\mathrm{old}} + \Delta\bm{S}_0)^{-1} \bm{m}_t^{\mathrm{old}}, \\
    \mathcal{L}_2 &= \frac{1}{2}\log|\bm{S}_0^\mathrm{new}| - \frac{1}{2} \log|\Delta\bm{S}_0| + \frac{1}{2} \log \left| \bm{S}_t^{\mathrm{old}} + \Delta\bm{S}_0 \right|
  \end{aligned}
\end{equation}

A critical issue in the gradient descent update occurs if $\bm{S}_0^{\mathrm{new}}$ is set to be identical to $\bm{S}_0^{\mathrm{old}}$. In this circumstance, $\Delta\bm{S}_0 = 0$, and the loss function becomes constant, which leads to vanishing gradients and prevents further learning. To address this issue, we apply the Woodbury matrix identity and properties of determinants to derive a reformulation of $\mathcal{L}_2$ that retains dependence on $\bm{S}_0^{\mathrm{new}}$, even when it is initialized as $\bm{S}_0^{\mathrm{old}}$:
\begin{align}
    \mathcal{L}_1 &= (\bm{m}_t^{\mathrm{old}})^{T} (\bm{S}_t^{\mathrm{old}} + \Delta\bm{S}_0)^{-1} \bm{m}_t^{\mathrm{old}}, \\
    \mathcal{L}_2 &= \log \left| \bm{S}_0^{\mathrm{new}} + \left[ \bm{I}_{n_\theta} - \bm{S}_0^{\mathrm{new}} (\bm{S}_0^{\mathrm{old}})^{-1} \right] \bm{S}_t^{\mathrm{old}} \right|
\end{align}

The derivation proceeds as follows:
\begin{itemize}
    \item The term $\mathcal{L}_1$ originates from the quadratic form of the Gaussian log-likelihood, where the covariance matrix is augmented to incorporate the increment $\Delta\bm{S}_0$.
    \item The term $\mathcal{L}_2$ results from applying the matrix determinant lemma (a direct consequence of the Woodbury identity) to the log-determinant component, allowing the determinant to be expressed in terms of $\bm{S}_0^{\mathrm{new}}$ and $\bm{S}_t^{\mathrm{old}}$.
\end{itemize}

In practical implementation, the optimization of $\bm{S}_0^{\mathrm{new}}$ is typically carried out iteratively. The procedure starts by initializing $\bm{S}_0^{\mathrm{new}}$ with $\bm{S}_0^{\mathrm{old}}$, after which the parameters of $\bm{S}_0^{\mathrm{new}}$ are updated at each iteration via gradient descent on the composite objective $\mathcal{L} = \mathcal{L}_1 + \mathcal{L}_2$. The closed-form expressions of $\mathcal{L}_1$ and $\mathcal{L}_2$ enable efficient gradient computation and ensure numerical robustness by preventing ill-conditioned or degenerate covariance updates.

\bibliography{reference}

\end{document}